\documentclass[]{spie}  %>>> use for US letter paper
\usepackage{amsmath,amsfonts,amssymb}
\usepackage{graphicx}
\usepackage{siunitx}
\usepackage[colorlinks=true, allcolors=blue]{hyperref}
\usepackage{lipsum} %remove after format is set.
\usepackage{multirow}
\usepackage{caption}
\usepackage{xcolor}
\usepackage[perpage]{footmisc} % counter for footpage
\usepackage[acronym]{glossaries} % for the newacronym command - added 201209
\usepackage{hyperref}

\newacronym{cmb}{CMB}{Cosmic Microwave Background}
\newacronym{vj}{VJ}{vacuum jacket}

\title{The design of the Ali CMB Polarization Telescope receiver}

\author[a,b]{M. Salatino}
\author[c]{J.E. Austermann}
\author[a,b]{K.L. Thompson}
\author[d]{P.A.R. Ade}
\author[a,b]{X. Bai}
\author[c]{J.A. Beall}
\author[c]{D.T. Becker}
\author[e]{Y. Cai}
\author[f]{Z. Chang}
\author[g]{D. Chen}
\author[h]{P. Chen}
\author[c,i]{J. Connors}
\author[j,k,e]{J. Delabrouille} % CNRS and IRFU/CEA and USTC
\author[c]{B. Dober}
\author[c]{S.M. Duff}
\author[f]{G. Gao}
\author[e]{S. Ghosh}
\author[a,b]{R.C. Givhan}
\author[c]{G.C. Hilton}
\author[l]{B. Hu}
\author[c]{J. Hubmayr}
\author[a,b]{E.D. Karpel}
\author[a,b]{C.-L. Kuo}
\author[f]{H. Li}
\author[e]{M. Li}
\author[f]{S.-Y. Li}
\author[f]{X. Li}
\author[f]{Y. Li}
\author[c]{M. Link}
\author[f,m]{H. Liu}
\author[g]{L. Liu} 
\author[f]{Y. Liu}
\author[f]{F. Lu}
\author[f]{X. Lu}
\author[c]{T. Lukas}

\author[c]{J.A.B. Mates}
\author[n]{J. Mathewson}
\author[n]{P. Mauskopf}
\author[n]{J. Meinke}
\author[a,b]{J.A. Montana-Lopez}
\author[n]{J. Moore}

\author[f]{J. Shi}
\author[n]{A.K. Sinclair}
\author[n]{R. Stephenson}
\author[h]{W. Sun}
\author[h]{Y.‐H. Tseng}
\author[d]{C. Tucker}

\author[c]{J.N. Ullom}
\author[c]{L.R. Vale}
\author[c]{J. van Lanen}
\author[c]{M.R. Vissers}

\author[c,i]{S. Walker}

\author[e]{B. Wang}

\author[f]{G. Wang}
\author[o]{J. Wang} 
\author[n]{E. Weeks}
\author[f]{D. Wu}
\author[a,b]{Y.-H. Wu}
\author[l]{J. Xia} 
\author[f]{H. Xu}

\author[o]{J. Yao} 
\author[g]{Y. Yao} 
\author[a,b]{K.W. Yoon}
\author[g]{B. Yue} %%
\author[f]{H. Zhai}
\author[f]{A. Zhang}
\author[f]{Laiyu Zhang}
\author[o,p]{Le Zhang} 
\author[o]{P. Zhang} 
\author[f]{T. Zhang}
\author[f]{Xinmin Zhang}
\author[f]{Yifei Zhang}
\author[f]{Yongjie Zhang}
\author[g]{G.-B. Zhao} %%
\author[e]{W. Zhao}

\affil[a]{Stanford University, Stanford, CA 94305, USA}
\affil[b]{Kavli Institute for Particle Astrophysics and Cosmology, Stanford, CA 94305, USA}
\affil[c]{National Institute of Standards and Technology, Boulder, CO 80305, USA}
\affil[d]{Cardiff University, Cardiff CF24 3AA, United Kingdom}
\affil[e]{University of Science and Technology of China, Hefei 230026} 
\affil[f]{Institute of High Energy Physics, Chinese Academy of Sciences, Beijing 100049}
\affil[g]{National Astronomical Observatories, Chinese Academy of Sciences, Beijing 100012}
\affil[h]{National Taiwan University, Taipei 10617}
\affil[i]{University of Colorado Boulder, Boulder, CO 80309, USA}
\affil[j]{IN2P3, CNRS, Laboratoire APC, Université de Paris, 75013 Paris, France} 
\affil[k]{IRFU, CEA, Universit\'e Paris-Saclay, 91191 Gif-sur-Yvette, France} 
\affil[l]{Beijing Normal University, Beijing 100875}
\affil[m]{Anhui University, Hefei 230039}
\affil[n]{Arizona  State  University, Tempe, AZ 85004, USA}
\affil[o]{Shanghai Jiao Tong University, Shanghai 200240}
\affil[p]{Sun Yat-Sen University, Zhuhai 519082}

\authorinfo{Send correspondence to Maria Salatino. E-mail: marias5@stanford.edu.}

\begin{document}
\maketitle
\begin{abstract}
Ali CMB Polarization Telescope (AliCPT-1) is the first CMB degree-scale polarimeter to be deployed on the Tibetan plateau at 5,250\,m above sea level. AliCPT-1 is a 90/150\,GHz 72\,cm aperture, two-lens refracting telescope cooled down to 4\,K. Alumina lenses, 800\,mm in diameter, image the CMB in a 33.4$^{\circ}$ field of view on a 636\,mm wide focal plane. The modularized focal plane consists of dichroic polarization-sensitive Transition-Edge Sensors (TESes). 
Each module includes 1,704 optically active TESes fabricated on a 150\,mm diameter silicon wafer. Each TES array is read out with a microwave multiplexing readout system capable of a multiplexing factor up to 2,048. Such a large multiplexing factor has allowed the practical deployment of tens of thousands of detectors, enabling the design of a receiver that can operate up to 19 TES arrays for a total of 32,376 TESes. AliCPT-1 leverages the technological advancements in the detector design from multiple generations of previously successful feedhorn-coupled polarimeters, and in the instrument design from BICEP-3, but applied on a larger scale. The cryostat receiver is currently under integration and testing. During the first deployment year, the focal plane will be populated with up to 4 TES arrays. Further TES arrays will be deployed in the following years, fully populating the focal plane with 19 arrays on the fourth deployment year. Here we present the AliCPT-1 receiver design, and how the design has been optimized to meet the experimental requirements.

\end{abstract}


\section{Introduction}\label{sec:intro}
The Cosmic Microwave Background (CMB), a relic radiation emitted when nuclei and electrons first combined to form neutral atoms in the Universe  at the ``last scattering'' epoch, when the Universe was about 380,000 years old, has been instrumental in establishing the standard cosmological scenario, the $\Lambda$CDM model. 
%% should we define what lambdaCDM means? 
Small amplitude anisotropies in the temperature of the background, witnesses of the primordial perturbations of the space-time metric that gave rise to present large-scale structures, convey essential information about physical processes at work in the Universe's infancy. The recent results of the Planck mission\cite{2011A&A...536A...1P,2020A&A...641A...1P} constrain with unprecedented precision and accuracy the main parameters of this  cosmological concordance model \cite{2020A&A...641A...6P}.

The CMB carries an additional wealth of information about the early Universe in the statistical properties of its polarization anisotropies \cite{1996AnPhy.246...49K,1997PhRvD..55.7368K}. Polarization maps
%%%, which arise from local quadrupole anisotropies in the radiation field at last scattering \cite{1997NewA....2..323H}, 
%%% NOOOO!!!!! those local quadrupole anisotropies at last scattering ONLY generate E-modes; it is a *generic* polarization pattern that can be decomposed into E and B modes
can be uniquely decomposed into two complementary modes, the so-called $E$ and $B$ modes \cite{1997NewA....2..323H,1997PhRvD..55.1830Z}. $E$ modes in the CMB were first detected with the DASI interferometer \cite{2002Natur.420..772K}, and have been observed with increasing precision by numerous experiments since then \cite{2004Sci...306..836R,2006ApJ...647..813M,2009ApJ...692.1247P,2007ApJS..170..335P,2009ApJ...705..978B,2010ApJ...711.1123C,2014JCAP...10..007N,2015ApJ...805...36C}. Those observations of polarization $E$ modes, and the full-sky observations by the Planck space mission, contribute to support the standard $\Lambda$CDM cosmological model. $B$ modes cannot be generated solely by scalar (density) perturbations at last scattering; some of them arise from the distortion of original $E$ modes by gravitational lensing on the photon path from the last scattering surface to the observer \cite{1998PhRvD..58b3003Z,2002ApJ...574..566H,2006PhR...429....1L}, and have already been detected at various levels of significance by a few CMB experiments \cite{2014ApJ...794..171P,2013PhRvL.111n1301H,2016ApJ...833..228B}. Other $B$ mode power can be generated by foreground sources, primarily within our own galaxy (dust, and hot electrons within magnetic fields).    

The next challenge is to detect large scale excess $B$ mode power above that predictable level of lensing $B$ modes and foregrounds, which would be a smoking gun for the existence of primordial tensor perturbations of the space-time metric, i.e. primordial gravitational waves. The existence and amplitude of those is key to understanding the physics at work in the early universe, at energies close to the Planck scale, well beyond those that can be probed by human-made accelerators. In particular, different models of cosmic inflation predict various levels and statistical properties of primordial gravitational waves, translating in different observable levels and properties of CMB polarization $B$ modes \cite{2016ARA&A..54..227K,2018JCAP...04..016F,2020arXiv200812619T}. For this reason, the hunt for primordial CMB polarization $B$ modes is one of the prime science targets of many CMB polarization experiments currently in construction or being studied, either ground-based\cite{Harrington16,Galitzki18,S419,Hamilton20}, balloon-borne \cite{2016SPIE.9914E..1JG,2018JLTP..193.1112G,PdB12}, or space-borne \cite{2011JCAP...07..025K,2014JCAP...02..006A,2018JCAP...04..014D,2019JLTP..194..443H,2019arXiv190901591D}.

%% PERHAPS NOT REALLY NECESSARY...
%As of today, cosmic inflation is one of the leading theoretical candidates in explaining some of the current paradoxes in observational cosmology. Arguably, inflation explains the observed spatial flatness of the universe and its observed isotropy (ref), as well as the absence of monopoles in the observable universe (ref). Alternative inflationary models predict a wide range of possible amplitude for the primordial CMB polarization B-mode signal. Hence, the detection of those B modes would uniquely constrain models of cosmic inflation, and more generally physics at work at energy levels which cannot be reached in any other way.

%% PERHAPS NOT REALLY NECESSARY...
%Cosmic microwave photons, on their way to us from the last scattering surface, are slightly deflected by existing matter structures. This effect, called CMB-lensing, smoothes the original CMB anisotropies emitted at last scattering. It also slightly distorts CMB polarization patterns. In particular, it converts some of the primordial E modes into observed lensing B-modes, which must be distinguished from primordial B modes to constrain models of cosmic inflation. For a standard cosmological model, the expected level of lensing B modes can be predicted. Within the error of this prediction, any excess of polarisation B modes would be an unambiguous signature of primordial gravitational waves in the early universe.

As of today, primordial polarization $B$ modes still escape observational detection. 
%{\color{red} REPLACE THIS TEXT IN RED WITH THE ONE AT THE BOTTOM? In 2014, results from sensitive observations with the BICEP2 experiment generated an international buzz with the announcement of a detection, with an amplitude of a fraction of a $\mu$K \cite{2014PhRvL.112x1101B}, of a signal in the $B$ mode polarization power spectrum on scales matching those expected from and inflationary signal, and with a level of primordial tensor modes of the order of 20\% of that of primordial scalar modes (for a tensor-to scalar ratio of $r \simeq 0.2$). Unfortunately, a re-analysis in conjunction with observations from the Planck mission concluded that the observed fluctuations of $B$ mode polarization were mostly due emission from interstellar dust grains in our own Milky way\cite{2015PhRvL.114j1301B}.} 
%JD: AGREED!
The BICEP2 observation of B-mode emission \cite{2014PhRvL.112x1101B}, and its re-analysis in conjunction with observations from the Planck mission\cite{2015PhRvL.114j1301B}, have underlined the importance of studying the emission from interstellar dust grains in our own Milky way. The current best upper limit for $r$, obtained combining together observations at 95, 150 and 220\,GHz, is $r_{0.05} < 0.07$ at 95\% confidence level\cite{BK15}.

The potential contamination of any future tentative detection of primordial $B$ modes in CMB polarization observations by emission from astrophysical foregrounds is a major concern. Emissions from the galactic interstellar medium, either through synchrotron emission from energetic electrons, or through thermal emission from elongated dust grains aligned by the galactic magnetic field, are particularly problematic. The combination of the two dominates the B-mode polarization in any region of sky, and at any frequency. This foreground emission hence has to be separated from CMB emission in the post-analysis of the observations. As the success of this task depends on the detailed properties of foreground emission, substantial effort is also ongoing to understand and model the polarized emission of galactic dust and synchrotron \cite{2013ApJ...765..159D,2013A&A...553A..96D,2013JCAP...06..041M,2016A&A...588A..65K,2017MNRAS.469.2821T,2017A&A...603A..62V,2018MNRAS.476.1310M,2020arXiv201102221K}.

The foreground problem has generated substantial literature in preparation of CMB $B$ mode observations \cite{2005MNRAS.360..935T,2009A&A...503..691B,2009AIPC.1141..222D,2016MNRAS.458.2032R,2017PhRvD..95d3504A,2018JCAP...04..023R,2018ApJ...853..127H,2019PhRvD..99d3529E}, with sometimes discrepant conclusions. In general, the strategy for minimizing foreground contamination in CMB map observations consists of observing the CMB in the cleanest possible regions of sky, at frequencies where the contamination by foreground emission is believed to be at a minimum relative to the CMB (around 60-100\,GHz), and to complement those observations with a detection of foregrounds at lower and higher frequencies, where synchrotron and dust emission from the galactic interstellar medium dominate \cite{2009LNP...665..159D}. However, from ground-based sites, the CMB can only be observed in specific windows of atmospheric transmission. Even in those, photon noise from the atmosphere, and emission from fluctuating inhomogeneities of water vapour, are a major source of noise in the observations \cite{2015ApJ...809...63E}. Hence, ground-based CMB experiments are preferably installed in cold and dry sites, such as the South Pole and the Atacama desert in Chile.
Existing experiments and those currently under development, such as Simons Observatory\cite{Galitzki18}, SPT-3G\cite{benson14}, BICEP3\cite{Ahmed14}, and BICEP Array\cite{Schillaci20}, will still only observe the sky from the Southern Hemisphere, as likely will the future CMB-S4\cite{S419}. Northern Hemisphere sites exist, and have hosted or host mm-wave polarization experiments such as AMIBA in Hawaii\cite{Wu09}, QUIJOTE in Tenerife \cite{2012SPIE.8444E..2YR}, or NIKA-2 in the Sierra Nevada in Southern Spain \cite{2014JLTP..176..787M}, but no kilo pixel focal plane camera has ever observed large patches of the Northern CMB sky in polarization. 

Recent studies of the MERRA-2\cite{Merra2} reanalysis\cite{Kuo17} and radiosondes from local weather station\cite{LiYP17,LiH18,LiH19} have demonstrated the suitability of the Ali site in Tibet for millimeter sky observations, confirmed also by an independent work\cite{Lapinov20} which has analyzed 12 years of the GEOS FP-IT data\cite{GMAO}. Ali1 and Ali2 sites are both located on the dry side of the Himalayas at 5,250 and about 6,100\,m above sea level, respectively. Ali1 is the selected site for the first 90 and 150\,GHz observations of the Chinese CMB observational program with the AliCPT-1 experiment\cite{LiH19}. With precipitable water vapour (PWV) quartiles equal to 0.871/1.343/2.125\,mm, Ali1 is an adequate site for millimeter and submillimeter sky observations.  PWV quartiles of 0.459/0.759/1.207\,mm make Ali2 very similar to Cerro Chajnantor in the Atacama desert\cite{Kuo17}.
With the rotation of the Earth, it is easy to observe a large fraction of the northern sky from the Ali site. During the primary observation season (from October to March), the average PWV is about 1\,mm and the atmospheric transmittance at millimeter wavelengths is high. Because of its location, 80°\,01$^\prime$\,E, 32°\,18$^\prime$\,N, Ali is a unique mid-latitude site for CMB observations in the Northern Hemisphere.

There are major advantages to complementing the South-Pole and Atacama observations with northern sky observations with comparable instrumental performance. This doubles the accessible low-foreground sky area, providing both redundancy in case of a tentative primordial B-mode detection, and reduced cosmic variance for the scientific exploitation of small-scale CMB polarization signals. It opens a new window for the joint exploitation of the observation of secondary CMB signals such as the Sunyaev Zel'dovich thermal and kinematic effects and CMB lensing signals with large scale structure surveys such as that of the DESI experiment, which is located at 31$^\circ$57$^\prime$ North \cite{2019BAAS...51g..57L}. Small-scale observations from a ground-based northern site will also improve the de-lensing capability that is key to detecting very low level primordial B-modes with a space mission targeting mostly the largest-scale B-modes, such as the JAXA LiteBIRD space mission \cite{2019JLTP..194..443H}.
%%% ahhh... but AliCPT-1 is a bit shy of aperture to get a good delensing map for LiteBIRD; only marginally better than LB itself (mid-frequency telescope aperture 400mm vs. AliCPT-1's 720mm, <2x)  

AliCPT-1 main science goals are the primordial CMB (primordial B modes, $r$), the measurement of the CMB lensing spectrum and production of CMB lensing potential maps, the search for cosmic birefringence\cite{Minami19}, the test for CPT symmetry\cite{SYLi15}, and the characterization of the properties of foreground emission in the northern sky. The first version of the AliCPT-1 instrument will observe in two frequency bands centered around 90 and 150\,GHz.  
Its angular resolution of 19$^\prime$ and 11$^\prime$ at 90 and 150\,GHz, respectively, is an intermediate resolution between that of the Small Aperture and Large Aperture telescopes of the Simons Observatory \cite{Galitzki18}. 
In this paper we present the design of the AliCPT-1 receiver, currently under integration and tests at Stanford University. Single pixel TES arrays have been designed and tested at NIST. The first TES array is currently being fabricated, while the corresponding feedhorn array is under testing at NIST. Arizona State University (ASU) is developing the RF cryogenic readout chain and room temperature readout electronics and associated software. The telescope mount has been deployed on the site and it is currently being commissioned. The expected first light is scheduled for 2021.

The paper is organized as follows. Section \ref{sec:receiver} outlines the entire receiver, with subsections including design drivers, optics, vacuum components, detectors, readout, and simulations performed to enhance the designs.  Section \ref{sec:site_mount} gives an overview of the site and the mount, followed by Section \ref{sec:conclusions} with conclusions.  

\medskip
%In this paper we present the design of the AliCPT receiver, the paper is organized as follows. Sec ... The receiver is currently under integration and tests at Stanford University. Single pixel TES arrays have been designed and tested at NIST. Efforts are on going to design the first TES array. The first feedhorn array is under test at NIST. ASU is developing the cold readout chain (LNAs and RF chain). The warm readout electronics is also under development at ASU. The telescope mount has been deployed on the site and it is currently being commissioned. The expected first light is scheduled for \textbf{202x}.

\section{Receiver design}\label{sec:receiver}
\subsection{Design Drivers} \label{sec:design drivers}

The driving function of the cryostat is to provide an optimal working environment for the core instrument: a two-lens refractive telescope and Transition-Edge Sensor (TES) detector array sensitive to the CMB polarization. The refractive
optics images the CMB on the focal plane populated by the detectors. Given the small amplitude of the target signal, $<$0.3\,$\mu$K, useful measurements necessitate a large number of photon-noise limited detectors and cold optical elements. This requires a suitable cryogenic vacuum housing able to maintain the detector arrays at $\sim$300\,mK. Moreover, it needs to maintain the aperture stop, the lens elements, and surrounding absorptive baffling at $\sim$4\,K in
order to minimize the instrument emission contribution to the detector signals.

\begin{figure}
\begin{center}
\includegraphics[width=0.75\textwidth]{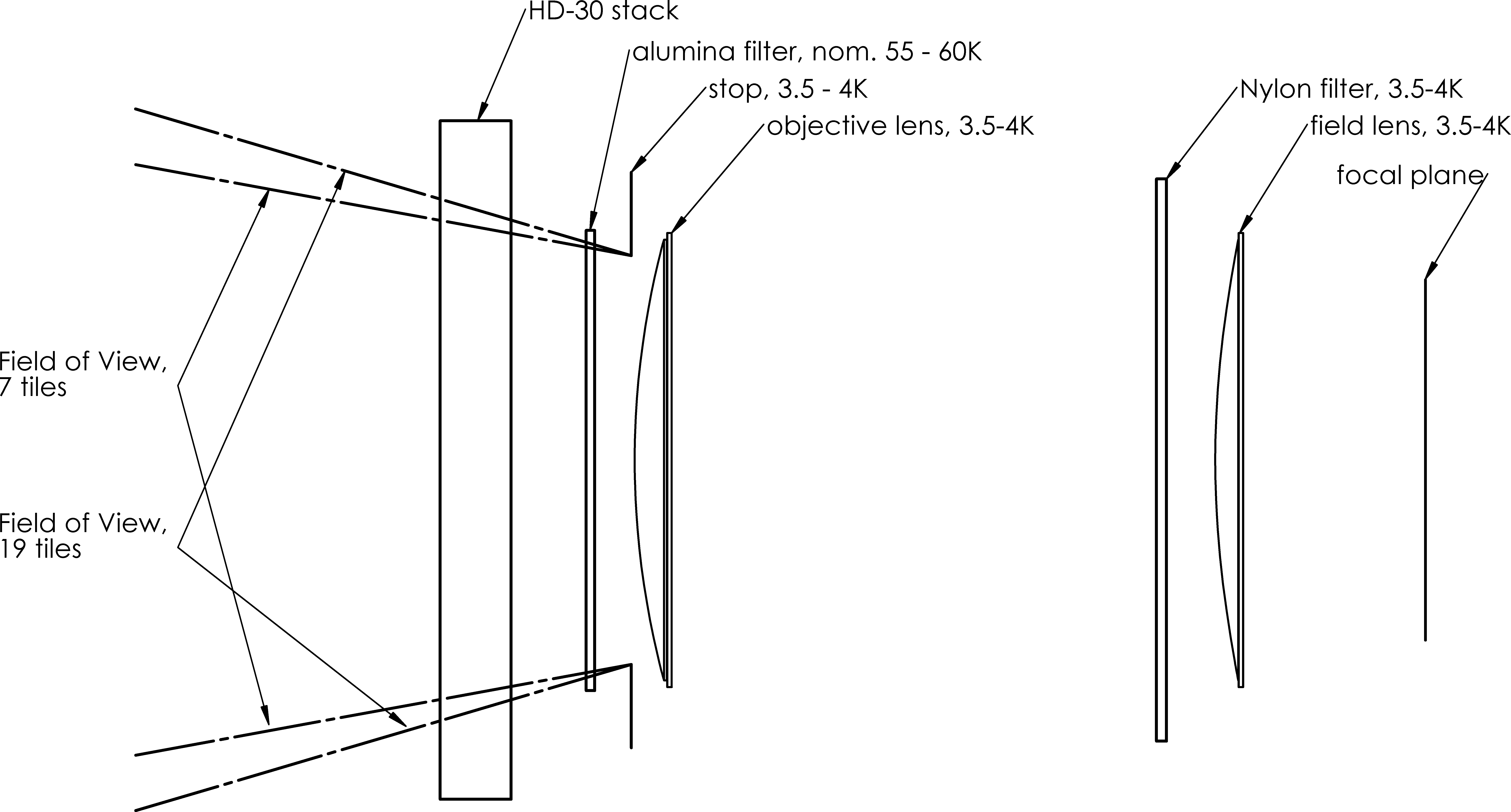}
\end{center}
\caption{The AliCPT-1 optical design. The differing FOVs for 7- and 19-module focal plane options are shown.  The stop is 720\,mm in diameter, and the alumina optics are each 800\,mm in diameter (filters are not to scale).  The f-ratio is 1.4 in the field center.  }\label{fig:optics_layout}
\end{figure}

The entire receiver is designed to operate with 19 detectors modules in a single AliCPT-1 cryostat.
However, the initial phase of AliCPT-1 will operate with a focal plane (FP) consisting of 4 detector array
modules, expandable up to 7 modules without changing the forebaffle design. A pair of 800\,mm diameter alumina ceramic lenses images the 20.8$^{\circ}$ Field Of View (FOV) onto the FP 379.6\,mm in
diameter in the case of up to 7 detector modules (Fig.\,\ref{fig:optics_layout}). The full 19-module receiver images the 33.4$^{\circ}$ FOV onto
a FP 635.6\,mm in diameter. The optics system volume, regardless of the number of detector modules, is 800\,mm diameter by 1348.2\,mm high. Together with the focal plane unit, this forms the fundamental instrument volume around which the rest of the cryostat is designed.
%, keeping with the critical requirements of 1,500 kg total weight (including forebaffle) and center-of-gravity within 200\,mm of the telescope mount elevation axis of rotation.

%\begin{figure}
%\begin{center}
%\includegraphics[width=0.45\textwidth]{figure/receiver/a1_chain.JPG}
%\end{center}
%\caption{The AliCPT-1 filter and electromagnetic band definition chain.}\label{fig:filter_chain}
%\end{figure}

\begin{figure}
\begin{center}
\includegraphics[width=1.\textwidth]{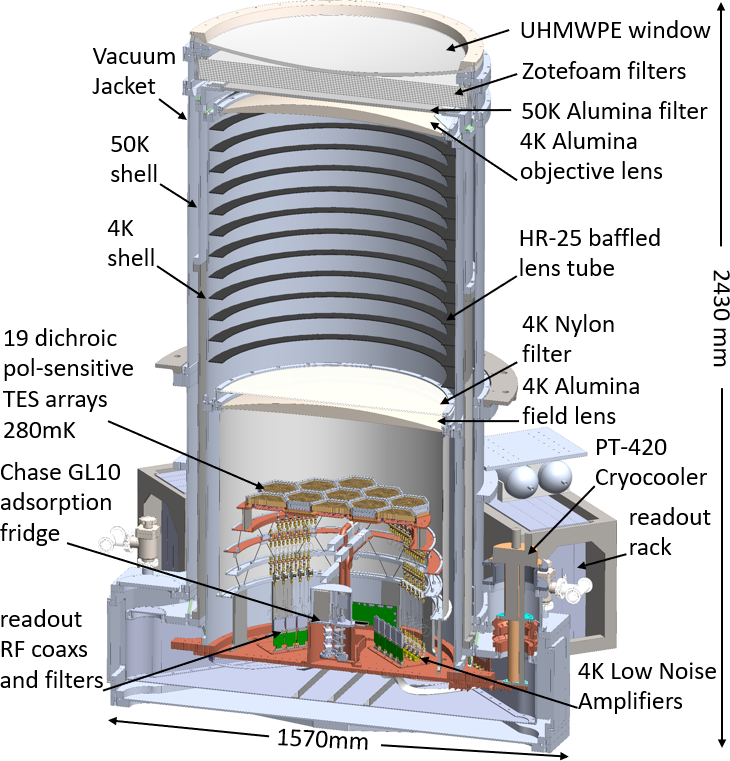}
\end{center}
\caption{Cross section of the AliCPT-1 receiver showing the main components.}\label{fig:receiver}
\end{figure}

\begin{table}[h]
\centering 
\begin{tabular}{|l|c|c|} 
\hline
\textbf{Frequencies} & \textbf{90GHz} & \textbf{150GHz} \\
\hline \hline
Center Frequency (GHz) & 91.4 & 145\\
Bandwidth (GHz) & 38 & 40 \\
%&& \\
%&& \\
Optical TES count & 16,188 & 16,188\\
P$_{sat}$ (pW) & 7.0  & 12.0\\
%Thermal Conductance, G$_{C}$ (pW/K)& 65.96 & 76.95\\
NEP Phonons (aW/$\sqrt{Hz}$)& 19 & 25 \\
NEP Photons (aW/$\sqrt{Hz}$)& 33 & 46 \\
NEP Total (aW/$\sqrt{Hz}$)& 38 & 53 \\
Resolution (FWHM) & 19' & 11'\\ 
NET CMB ($\mu K\sqrt{s})$ & 274 & 348 \\
\hline
\end{tabular}
\caption{AliCPT-1 per-detector specifications.  Total TES count is for the full 19 module configuration. 
TES parameters assume $T_{bath}=280\,$mK, $T_{c}=420\,$mK, band-weighted efficiency 73\%. 
Noise values are calculated assuming a median atmospheric emissivity for the site based on 3d MERRA2 data\cite{Merra2}, an atmosphere effective temperature of 258\,K and a 60$^{\circ}$ elevation of observation.  Saturation power, P$_{sat}$, is chosen such that the TES are modelled to saturate on the 95th-percentile weather conditions and observing at our lowest projected elevation of 28$^{\circ}$.
%G$_{C}$=Thermal Conductance, 
NEP=Noise Equivalent Power, 
NET= Noise Equivalent Temperature, 
FWHM=Telescope beam Full Width at Half Maximum projected on the sky.
}
\label{table:FPU specs}
\end{table}

\begin{figure}
\begin{center}
\includegraphics[width=1.\textwidth]{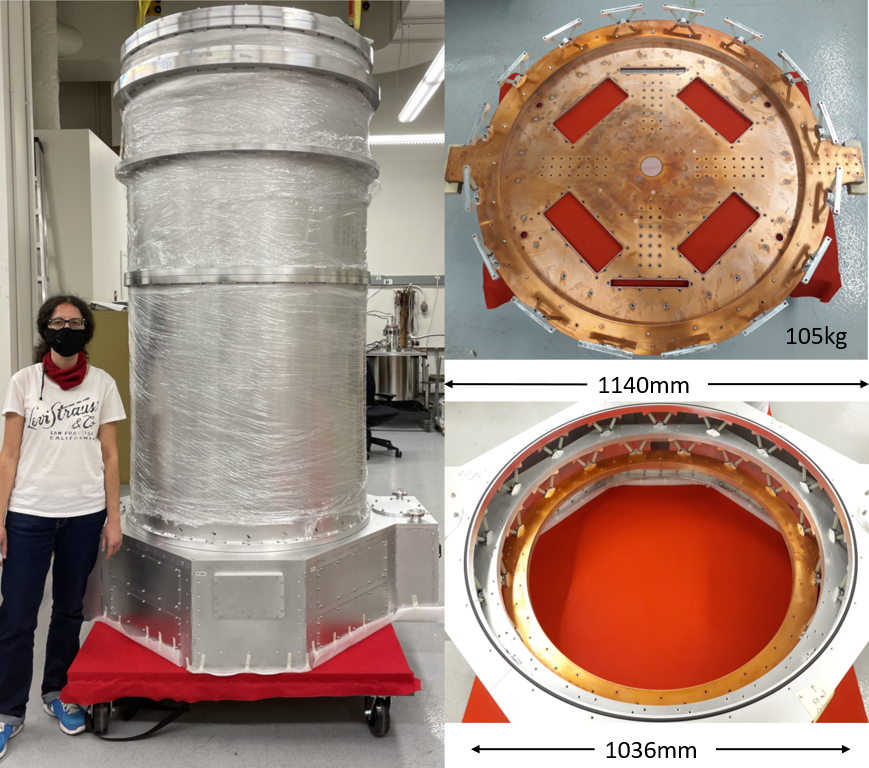}
\end{center}
\caption{Left: the AliCPT-1 300\,K VJ. Top right: 4\,K OFHC stage, bottom right: G-10CR truss assemblies across the 300\,K-4\,K temperature stages.}\label{fig:VJ}
\end{figure}

\subsection{Optical Design} \label{sec:opt_design}

The optical design of AliCPT-1 follows the form of the BICEP and Keck series of cryogenic refractors\cite{Ahmed14} (see Fig. \ref{fig:optics_layout}).  It has an f/1.4, 2-lens camera at~4\,K, infrared filters at~50\,K and~4\,K, and a microwave-transparent window.  The focal plane is telecentric, and the aperture stop is in front of the objective lens. The stop is 720\,mm in diameter. The lenses and the 50\,K filter are alumina, 800\,mm in diameter, manufactured by Kyocera\footnote{Kyocera International, Inc., San Diego, CA, USA.}.  The window (not shown) is ultra-high molecular weight polyethylene (UHMWPE), 22\,mm thick for sea level lab testing, and 12\,mm thick for use at the high altitude site.  From the stop through the field lens is mounted on the 4K stage.  

There are four stages of filtering.  The first stage is an infrared blocking stack of 12 Zotefoam HD-30\footnote{Zotefoams PLC, Croydon, U.K.} filters, each 3.2\,mm thick, mounted on the inside of the cryostat just below the window\cite{Choi13,Kang18}.  The second stage is the 10mm thick alumina filter, described above, mounted on the 50K stage.  The third stage is a Nylon filter, mounted on the 4K stage just upstream of the field lens. A fourth stage of low-pass edge filters will be mounted on each detector module (Sec.\,\ref{sec:det_module} and Fig.\,\ref{fig:det_module}).  

Microwave absorbing material is located on the inside of the cryostat at strategic locations along the optical path.  The baffles seen in fig.\,\ref{fig:VJ}\, are representative; they have not been finalized but are expected to follow a similar scheme as BICEP3, a corrugated structure made of Eccosorb HR-25\footnote{Laird Technologies, Inc., Earth City, MO, USA} with an epoxy matrix to stabilize it from cold vacuum degradation.  

The anti-reflection (AR) treatment will consist of 2 layers of Stycast\footnote{Henkel AG \& Co., Düsseldorf, Germany.} epoxy on each alumina component surface\cite{Inoue14,Ahmed14}.  We adopt a new scheme for epoxy AR, by casting flat sheets of precision thickness in a mold, softening the sheets in an oven at moderate temperatures to allow them to form to lens profiles (or left flat for filters), then bonding the epoxy layers to the optics with low density polyethylene (LDPE) film or similar low-temperature melt-point plastic using vacuum-bagging.  The AR treatments for the window and the Nylon filter have not been finalized, but both expanded polytetrafluoroethylene (PTFE) and metamaterial\cite{Raguin_Morris93,Biber03} (grooves or cross-cuts in a dielectric) are current options\footnote{Grooves or cuts in the high-stress window are dangerous, so the option being considered consists of machining these features in a thin layer of polyethylene that is then bonded to the window surfaces with LDPE film, slightly increasing the net window thickness but protecting the window from propagating cracks.}.  Our goals are to limit mean surface reflections in each band to under 1\% per surface.  

To accommodate field curvature, the outer tiles (12 each) will be slightly elevated with respect to the inner tiles (central tile and first ring of 6 tiles).  The resulting Strehl ratio at 150\,GHz of the design is $> 0.96$ for the inner tiles (radius  $< 200$\,mm) and $> 0.90$ for the outer tiles (radius $> 152$\,mm).  Strehl over large areas of the focal plane would be reduced with a best compromise focal position for all detectors in the same plane.  The lenses, however, are built to a looser specification than we originally requested (peak to valley deviation of the profile $< 0.05$\,mm vs. $<0.20$\,mm).  Metrology data inserted into the optics modeling program show Strehl for the inner (outer) tiles will be $> 0.94$ ($> 0.88$) for either of two combinations of the two objects and two field lenses manufactured.  They are, however, very good surfaces of revolution (RMS $\sim$2\,-\,6\,$\mu$m along a concentric circle), with deviations from the design profile only on large scales.  

Not shown in the above figures are the two options for forebaffles (7-tile version seen in fig.\,\ref{fig:mount}, below).  Models for forebaffles and groundshields made early in the design process indicated the groundshield would need to be impractically large, based on requiring at least 2 diffraction or surface scattering photon events to couple ground emitters (not including the reflective groundshield) into the receiver aperture.  We thus reworked the scheme to involve only a forebaffle, and permitted scattering from the edges of the physical baffle rings to couple with just a single event (representing a small fraction of the solid angle seen by the receiver).  However, the high winds at the site required the addition of the wind break, described below (Sec.\,\ref{sec:site_mount}).

\subsection{Mechanical and Thermal Cryostat Design} \label{sec:mech_design}

The cryostat receiver, hosted in a custom vacuum jacket (VJ) 1,570\,mm wide and 1,430\,mm tall (Fig.\,\ref{fig:receiver}), is designed to fit the AliCPT-1 mount (Sec.\,\ref{sec:site_mount}). The cryostat receiver, built by Atlas Technologies and Oakland Machine Works\footnote{\url{https://www.atlasuhv.com/}, \url{http://www.oaklandmachineworks.com/}.}, was designed and is currently being constructed and tested at Stanford University. 

The mechanical design was driven by easy access to all the subsystems and  assembly/disassembly of the cryostat. It has two main independent sections: the optics section and the camera section. The former includes a 50\,K and a 4\,K radiation shield nested inside the 300\,K VJ, a room temperature window, the cold lenses and the infrared-blocking filters. The lower section of the cryostat contains the detector arrays (Sec.\,\ref{sec:det_module}), mounted on a Focal Plane Unit (FPU) assembly\cite{Salatino20}, and the RF cryogenic readout chain (Sec.\,\ref{sec:readout_electronics}), all inside the 4\,K volume.  Also in the lower section are the two cooling systems. 
The 300\,K VJ and the 50\,K radiation shield support Zotefoam and Alumina filters, respectively. Finally, the 4\,K radiation shield supports the Alumina objective and field lenses and a Nylon filter.
The 300\,K, 50\,K and 4\,K stages are mechanically connected and thermally isolated from each other by two series of G-10CR bars and truss assemblies, at the top and at the bottom of the cryostat, respectively, similar to a design from BICEP3\cite{Ahmed14}. The G10-CR bars guarantee the axial alignment of the 4\,K optics tube with respect to the VJ and the telescope mount.  The 50K stage is covered by 30 layers of multi-layer insulation (MLI), and the 4K stage by 5 layers.  

Considerations of the operating environment resulted in two different access methods. The first is more traditional, in which the VJ tubes and window section, the 50\,K radiation shield tubes, and the 4\,K radiation shield tubes containing the lenses are removed, leaving the exposed FPU unit easily accessed from above. None of the thermal links nor any cables except for those servicing a few thermometers on the 50\,K and 4\,K shields need be detached. The second mode involves removing the bottom VJ cover and a cover on the bottom of the 50\,K radiation shield. Then the 4\,K baseplate, supporting the entire FPU and fridge, can be detached and lowered down. A substantial number of thermometry and readout cables need to be detached, along with one thermal link to the pulse tube second stage.  However, this bottom access method allows quick repair of readout and focal plane issues, and can be performed while the receiver is still in the mount.  

Side panels in the VJ bottom lateral surface provide ports to warm readout electronics and HouseKeeping (HK) (Sec.\,\ref{sec:HK}), as well as access to the cryocooler cold heads.  Electronics attached to the outside of those parts of the VJ handle detector readout and HK, and the digital data is communicated by those units to the control computer system at ground level for merging, reformatting, and storage.  

Cooling is provided by two systems. A Cryomech\footnote{Cryomech, Inc., Syracuse, NY, USA, \url{https://www.cryomech.com/products/pt420/}.} PT420 2-stage pulsetube cryocooler cools (nominal) 50\,K and 4\,K stages. The sub-Kelvin cooling is provided by a custom Chase\footnote{Chase Research Cryogenics, Sheffield, U.K., \url{https://www.chasecryogenics.com/}.} 3-stage adsorption fridge with a \textsuperscript{4}He head at $\sim$1\,K, a \textsuperscript{3}He head at $\sim$350\,mK, and a dual \textsuperscript{3}He head at $\sim$280\,mK. Oxygen-free copper heat links connect the cold heads to the relevant stages for both the pulsetube and the adsorption fridge, and isolate the vibration of the cryocooler from the rest of the cryostat. The FPU assembly and the sub-K adsorption fridge (Sec.\,\ref{sec:therm_perf}) are
mechanically mounted on an oxygen-free, high conductivity (OFHC) 25\,mm thick copper baseplate closely attached to the 2nd stage of the pulsetube cryocooler head. The total mass of the receiver is 1,630 and 2,010\,kg in the case of 7- and 19-detector modules, respectively. These estimates include forebaffles of 150 and 460\,kg, respectively.

\subsubsection{Predicted Mechanical Performance} \label{sec:mech perf}
The design of the cryostat for the 7-detector module and one PT420 assembly configuration has been optimized with a thermo-mechanical Finite Element Analysis (FEA) with COMSOL\footnote{COMSOL Multiphysics software:  \url{https://www.comsol.com/}.}. 
The geometric model includes all the major parts of the receiver, has a total weight of about 1,500\,kg and has been meshed with $\sim2.3\cdot10^6$ tetrahedral elements with overall mesh quality equal to 0.62.
%and the performance has been tested

%The geometric model includes all the radiation shields and baserings, G10-CR trusses and bars, the FPU 4K ring and 4K supports, placeholders for the sub-K assembly and the TES arrays, filters and mounting supports, a window assembly placeholder, the HK and the readout racks and one PT-420 assembly. The total weight of the geometric model is about 1,500\,kg. The geometric model.

We have studied the configuration that maximizes the static mechanical stress, i.e. receiver on the mount at 45$^{\circ}$ elevation. Thermal contraction and  conductive heat trasfer of all the involved parts have been modeled using materials data from COMSOL libraries and the NIST Cryogenic Material Properties Database \cite{NISTdatabase}.

The top surface of the 50\,K and 4\,K baserings have been set to be 39\,K and 3.55\,K, respectively, the nominal, expected temperatures (Sec.\,\ref{sec:therm_perf}). We have applied the expected radiative thermal loads
on the 50\,K alumina filter, 50\,K tube and 4\,K objective lenses 
equal to 25.5\,W, 5\,W and 250\,mW, respectively. %(we have considered ''warp” direction for the G-10CR). 

The majority of the mechanical stress arises from the Al differential thermal contraction in the (300-50)\,K temperature range: since the 50\,K tube is 1,257\,mm\,tall it contracts by 7.3\,mm once cooled down to 50\,K, in addition to the radial contraction; this results in a mechanical displacement vector relative to the top attachment point inside the cryostat wall pointing at a 14$^\circ$ angle with respect to the tube axis.
To reduce the resulting mechanical stress experienced by the VJ-50\,K G-10\,CR bars, they are mounted at the same tilt angle mentioned above so their flattened geometry is favorable for the flex direction.

The mechanical design of the G-10\,CR truss assembly has been optimized while keeping the same thermal load. We have achieved a final improvement over a BICEP3 style truss for maximum stress equal to 369MPa/279MPa = 1.32, with a displacement induced by differential thermal contraction that is 40\,\% larger than the one experienced by BICEP3.

%A BICEP3 design style for the VJ-50K trusses has been investigated with a thermo-mechanical FEA of the AliCPT-1 VJ and 50K stages placeholders (i.e. two Al plates 2000mm diameter, 100mm thick, 843kg). In BICEP3, the G-10CR trusses rods showed a maximum von Mises stress and displacement equal to 369\,MPa and 1.28\,mm, respectively. Adopting the BICEP3 truss design for AliCPT-1 would almost double the stress (660\,MPa) being 2.1\,mm the expected displacement. By replacing a single BICEP3 rod (7.9\,mm diameter) with 4 ones 4\,mm diameter keeping constant the estimated truss conductive thermal load (i.e. the $A/l$ ratio where $A$ and $l$ are the rod area and length, respectively) we have achieved a final improvement over BICEP3 for maximum stress equal to 369MPa/279MPa = 1.32.

%The VJ base bottom cover should maintain the high vacuum inside the cryostat and to withstand the stress induced by the pressure difference between the atmosphere outside and the high vacuum inside the cryostat with a safety factor above 3 and a maximum deflection of the cover under 10\,mm (to prevent damaging the equipment inside the VJ). An Al plate 29\,mm thick ensures a safety factor of 3, 23 mm deep cut-out features reduce the weight while maintaining the mechanical strength. 
Design simulation has allowed reducing the mass of the VJ base bottom cover from 100.8 kg to 56.5 kg.  Patterned cutouts aligned with the deflection gradient of the cover retain most of its stiffness and strength while removing a significant fraction of the material volume compared to a solid plate (Fig.\,\ref{fig:fpu_fea}, left). The plate is left at full thickness near the o-ring area, identified to be a weak spot. Mechanical FEA demonstrated a safety factor of 3.02, and the maximum deflection of 3\,mm.

\subsubsection{Predicted Thermal Performance} \label{sec:therm_perf}
The first phase of AliCPT-1 will be cooled by a single PT420. However, the entire receiver design is fully compatible with the installation of a second PT420, in case it is necessary for future upgrades beyond 7-detector modules. The AliCPT-1 thermal design is driven by the need to minimize the radiative loading of the FP and the detectors and being able to cycle the fridge. 

%{\color{blue}The uncertainty in whether or not a second PT420 will be needed is driven primarily by uncertainty in the 4K power dissipation of the LNAs and associated electronics.}
%These drivers have been translated into goal base temperatures for various parts of the cryostat. These goals are  guidelines, not hard design constraints, and we aim to achieve even better performance.  

%In order to translate temperatures into part design constraints we calculate or simulate the temperature difference, $\Delta T$, across the part while it is loaded, and then add up these values from the expected base temperature of the PT420's cold heads to estimate the temperature of the critical part.

In order to cycle the fridge, the fridge baseplate must be below 5.2K while the fridge is loading the 4K stage with 1\,W of power, roughly the average power output of the fridge at the end of Helium condensation.
%(even if significant performance is gained all the way to 3.3K)
We target to keep the top of 4\,K optics tube below 4.5\,K, preferably below 4\,K: this temperature directly effects the excess radiation both on the detectors and the FP. We aim to keep the 50\,K alumina filter below 60\,K, preferably closer to 55\,K since this temperature will dominate the radiative load onto the 4\,K optics tube. Below we report the predicted thermal performance.

%\textbf{Thermal interfaces}.
%Based on the Stanford team expertise, gained during the BICEP-3 experiment, we are confident in achieving good thermal joints with combinations of n-grease, gold plated joints and high clamping force and to reliably make metal to metal joints at 4K with conductances of 100~W/K. 

%% suggested text to replace the following named paragraphs:

We have modeled the critical 50\,K and 4\,K stage thermal performance (Tab.\,\ref{table:thermal_loads}). The inputs are the predicted aperture loading through the IR filters and the heat generated on 4\,K during the fridge cycles, and the radiative and conductive loading from the vacuum jacket.  Basing performance on the IR filtering and thermal performance of the metal-metal and metal-optics interfaces of BICEP3, we have estimated total loading and thermal gradients acceptable to keep the 50\,K filter, the 4\,K stop and lenses, and the fridge heat sink to acceptable temperature levels during operation and fridge cycling based on the single PT-420 cryocooler, with margin.

\begin{table}[h]
\centering 
\begin{tabular}{|l|c|c|c|c|c|} 
\hline
\textbf{Power source}\ & \textbf{50K\,(W)} & \textbf{4K\,(mW)}  & \textbf{1K\,($\mu$W)}  & \textbf{350mK\,($\mu$W)} & \textbf{280mK\,($\mu$W)}\\ 
\hline \hline
Aperture radiative & 25.5 & 250 & 6.5& 0.03& 1.5-4.0\\
G-10 CR bars & 0.8 & 60 & - & - & - \\
G-10 CR trusses & 4.3 & 180 & - & - & -\\
carbon fiber trusses & - & - & 63.0 & 4.1 & 0.2\\
Non-aperture radiative & 5.0 & 10 & - & - & - \\
LNAs (upper limit) & 1.1 - 2.9 & 70 - 190 & - & - & -\\
DC cabling & 0.6 - 1.5 & 5.7 - 15.4 & 25.1 - 65.5 & 1.5 - 4.1 & 0.2 - 0.5\\
RF cabling & 0.8 - 2.3 & 37 - 101 & 7.6 - 20.5 & 0.2 - 0.6 & 0.7 - 1.9\\
Readout attenuator & -  & 1.1 - 2.9 & 1.4 - 3.8 & 1.9 - 5.1 & 0.6-1.7\\
Temperature control & - & - & - & - & 3.6\\
TES bias & - & - & - & - & 1.4 - 3.8\\
\hline
Projected Total Loading& \textbf{38.1 - 42.3} & \textbf{614.0 - 809.3} & \textbf{103.6 - 159.3} & \textbf{7.8 - 15.7} & \textbf{8.2 - 14.0} \\
\hline
Nominal Cooling Power & \textbf{47.3} & \textbf{1000} & \textbf{150} & \textbf{30} & \textbf{20}\\
\hline
\end{tabular}
\caption{Thermal budget with ranges representing the 7 and 19-detector module configurations. The listed available cooling power at 50K and 4K is from one single PT420. LNA= Low Noise Amplifier.} 
\label{table:thermal_loads}
\end{table}

Given the uncertainties, it is possible that insufficient cooling power will be available during fridge cycles to maximize helium condensation in the adsorption fridge, in order to ensure a one day observing cadence with the full 19-tile configuration including all the power dissipating amplifiers. In this event, we may install a second PT-420 cryocooler when we upgrade to a fully populated focal plane.

We have also modeled the sub-Kelvin stages, based on radiative loading, readout power dissipation, conductive loads, and a small additional power for temperature control (Tab.\,\ref{table:thermal_loads}). We estimate the detectors can operate with a base temperature $<$280\,mK and with a hold time of $>$20 hours.  While the margins are slim, we have several options in case of sub-optimal performance, such as increasing the number of Zotefoam IR filter layers or increasing the thickness of the Nylon filter (or adding a second), both of which should decrease radiative loading of the FPU (with minor performance increase and minor disadvantages), all the way to acquiring a larger capacity fridge and/or adding the second PT-420 cryocooler.

\subsection{Magnetic shielding}
%TES readout makes extensive use of SQUIDs, because their resistance, in the $\Omega$ range, matches the TES's one and their easiness to be multiplexed. However, their sensitivity to magnetic shields (e.g. Earth, electronic devices, wifi, etc. etc.) 
The SQUIDs require optimal shields to prevent spurious effects on the TES output signal (Sec.\,\ref{sec:dets_readout}). The best shielding strategy arises from combining together different shielding materials at different temperatures. A  645\,mm tall Nb cylindrical shield, fully enclosing the FPU, will be mounted on the FPU 4\,K ring. 
An Amuneal magnetic shield\footnote{Amuneal Manufacturing Corp.: \url{https://www.amuneal.com/}.} will be mounted on the outer walls of the 50\,K tube, made of two overlapping rolled sheets, not welded into fully closed tubes so that they can be fitted over the external flanges on the 50\,K shield. The gaps in the two layers will be offset by 90$^{\circ}$, nearly eliminating the degradation from the gaps, according to the FEA simulations used to optimize the design. In addition, a layer of Nb will be deposited or added between the detectors and readout components. 
If additional shielding proves necessary, it is possible to add additional Nb shielding on the detector module (Sec.\,\ref{sec:det_module}).

\subsection{The Focal Plane Unit}\label{sec:FPU}
\begin{figure}
\begin{center}
\includegraphics[width=1.\textwidth]{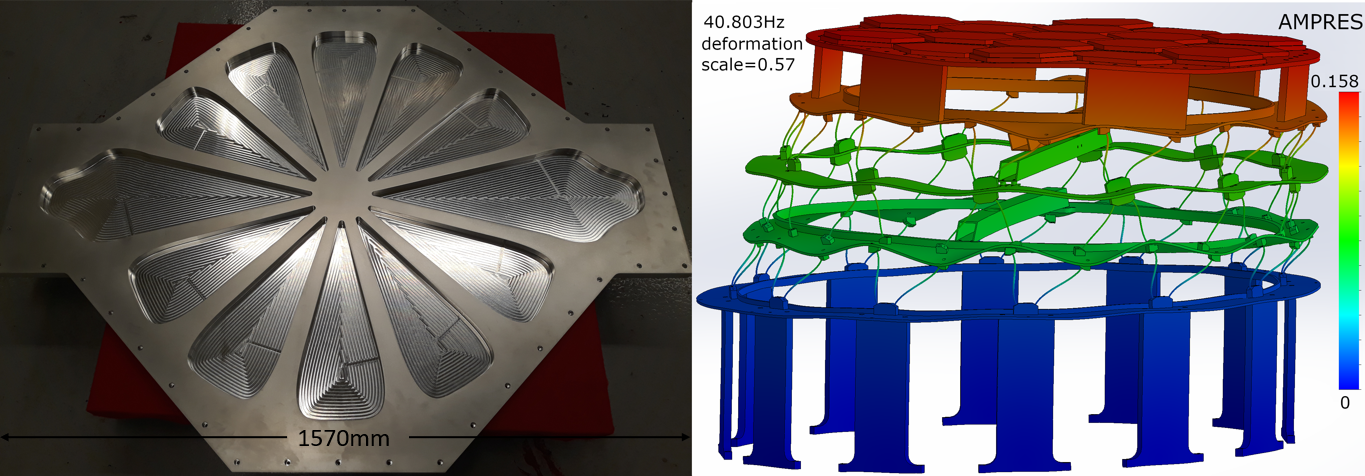}
\end{center}
\caption{Left: VJ bottom cover. Right: resonance frequency study of the FPU; the first resonance mode, parallel to the planes of the FPU rings, is 40.8\,Hz.}\label{fig:fpu_fea}
\end{figure}

A Focal Plane Unit (FPU), with three isolated thermal stages each cooled by one of the fridge cold heads (described above), mechanically supports the 19 detector modules. The FPU thermo-mechanical design is described in\cite{Salatino20}, here we focus on the design optimization. The structural supports are low thermal conductivity carbon fiber reinforced plastic (CFRP) trusses, following a design already implemented in BICEP3\cite{Ahmed14}, with heritage in the earlier BICEP, Keck, and SPIDER receivers, as well as other instruments in the field. CFRP provides adequate stiffness while minimizing the thermal load\cite{RunyanJones2008}. The FPU mechanical design has been optimized with thermo-mechanical FEA simulations, with 19 detector modules and at 45\textdegree\, elevation, generating the maximum static shear stress. The maximum von Mises stress experienced by the truss assemblies is $\sim$30\,MPa, far below their tensile strength of 1.72\,GPa\footnote{CST The Composite Store, Inc.: \url{http://www.cstsales.com/}.}. The FPU requires resonance frequencies well above 10\,Hz to avoid the low frequency pulsetube excitation from leaking into the detector signal. The introduction of a 15\textdegree\, angular offset in the angular position of the 1\,K-4\,K trusses, for example, increased the first two resonance modes, oscillations in plane parallel with the FPU rings, by 39\%  (Fig.\,\ref{fig:fpu_fea}, right). The first three resonance modes in the final design are at 51.8, 52.5, and 77.7\,Hz (the third is a drum mode of the detector mounting plate) and 40.8, 41.3, and 86.2\,Hz for the 7 and 19-detector modules, respectively. Simulations included mock masses representing the detector modules, but only some components of the three fridge heat straps and no electronic cables or their heatsinks, so these simulations are only approximations of the final performance.

%The TES arrays are cooled down to about 280\,mK by a custom Chase GL10 adsorption fridge. The 20\,$\mu$W UC head cooling power is achieved by two equal cold heads running in series.

Temperature-control modules control the temperature of both the FP and a plate heatsunk to the ultra cold (UC) head, using SRS SIM 960 Analog PID temperature controller electronics units outside the cryostat. Stainless steel thermal filters between the FP and the UC isolate the FP from thermal fluctuations of the UC head.

%%Suitable Cu plates heatsink the HK \ref{sec:HK} and readout \ref{sec:cold readout} cables at each FPU temperature stage. The FPU is RF-shielded by a thin Aluminized film, following a design implemented in the SPT camera.

\subsection{Signal chain}

\subsubsection{Detectors and multiplexing}
The TESes will be read out by a $\mu$mux readout architecture where each TES is inductively coupled to a RF-squid which is, in turn, connected to a resonator with unique resonance frequency\cite{Irwin04,Mates11,Dober17}. In a $\mu$mux architecture, the absorption of CMB photons ultimately result in a shift of the resonant frequency.
For the initial deployment phase, the detectors of a single module will be split into two resonator networks, each operating in the 4--6\,GHz range. An ongoing development effort aims to eventually operate the entire module of detectors on a single network spanning 4--8~GHz. 
%Difficulties in doubling the resonators range reside in unexpected losses and challenges in achieving a sufficiently high resonator quality factor. 
Resonator pre-screening before deployment will help ensure optimal resonator performance and frequency spacing to minimize resonator collisions and crosstalk.   

The AliCPT-1 FP will be populated by 19 arrays of polarization-sensitive AlMn dichroic TES bolometers, fabricated on 150\,mm diameter wafers, and incorporating a strong technological heritage of OMT-coupled dichroic detectors\cite{duff16}, material properties matched to a $\sim280$\,mK bath temperature\cite{hubmayr16}, lumped-element microwave termination\cite{walker20a}, and in-series Al TES for in-lab characterization\cite{hubmayr16}. When fully populated with 19 modules, AliCPT-1 will comprise 32,376 optical detectors equally split between the 90 and 150\,GHz bands. Each array pixel comprises four TESes, each one sensitive to one linear polarization and one frequency band. Each array consists of 426 feedhorn coupled and optically active pixels and 4 optically ``dark" pixels used for diagnostic and characterization purposes. The first detector array is under fabrication, while measurements of single pixel prototypes are presented in \cite{Salatino20, Walker20}.

\label{sec:dets_readout}

\subsubsection{Detector module}
%The TESes are optically coupled to the incoming radiation through a Si feedhorn array, built using a technology already developed for AdvACT: a series of Si platelets with suitable holes, each one represents a section of the horn profile, are aligned and glued together. The detectors will be fabricated on 150\,mm diameter Silicon wafers at NIST, Boulder. Each wafer will contain 1,704 optically active TESes. Mn concentration will be suitably doped to tune the $T_c$ of the Al superconducting film, following a well established technology used for AdvACT \cite{Li16}. Since the TES arrays will be cooled down to 280\,mK with the adsorption fridge, the $T_c$ is tuned to be 420\,mK.

%Each TES array will be hosted in a hex-shaped detector module which provides the mechanical support and the thermal contact to the detectors. The detector module contains the resonators chips, connectors for both the RF and the DC chain. Springs are present to assure the thermal contact between the Si wafer and the Cu part. A Nb shield provides the first stage of magnetic shielding to the detectors and resonators chips. All the wire bonds, interconnects between the TESes, the resonator chips and the input/output connectors are self-contained in the hex-shaped module, allowing an easy integration on the FPU. 
%The FPU is designed to be populated by up to 19 detector arrays.
%AliCPT will follow a staged deployment. From one to four TES arrays will be deployed on the first phase, up to ten for the second phase. Finally,  fifteen to nineteen arrays will be deployed between the third and forth phase.

\begin{figure}
\begin{center}
\includegraphics[width=0.85\textwidth]{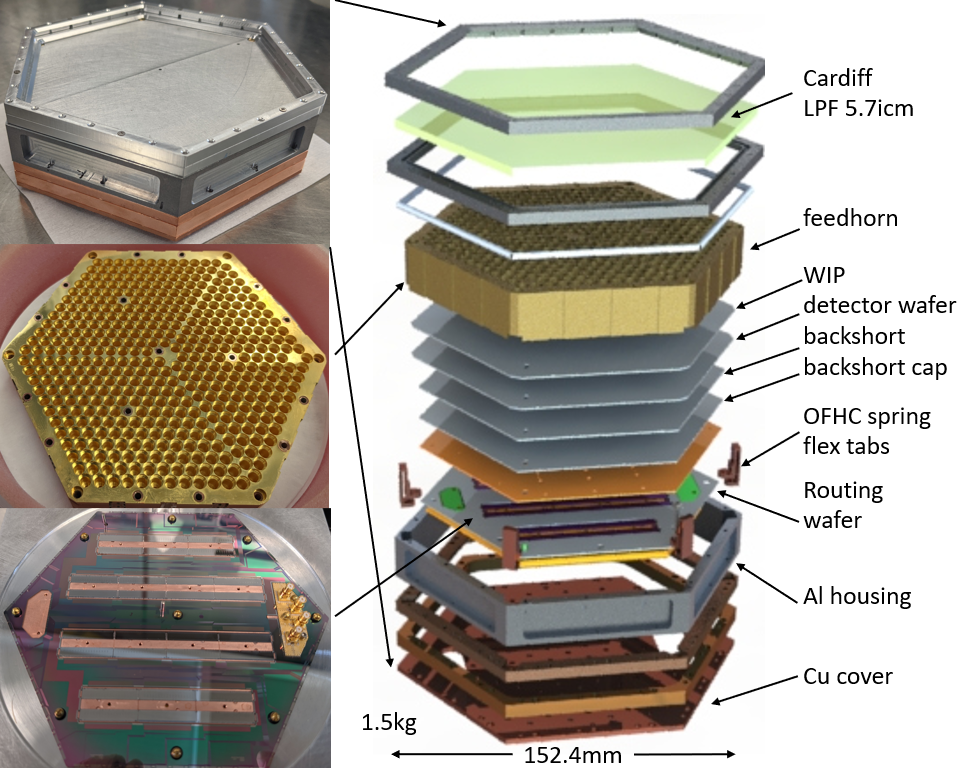}
\end{center}
\caption{AliCPT-1 detector module. Left: real components. Right: SolidWorks exploded view.}\label{fig:det_module}
\end{figure}

The 636\,mm diameter FP will be fully-filled by 
19 independent detector modules (Fig.~\ref{fig:det_module}) tiled in a hexagonal pattern.  The footprint of each module is a hexagon with a circumcircle diameter of 152.4\,mm (132\,mm between parallel sides of hexagon). 
%The module consists primarily of optical coupling components (feedhorns, waveguide, backshort), the dichroic detector array, $\mu$mux and bias electronics, a free-space low-pass optical filter, and the mechanical housing.  
The detector module optically couples the incoming polarized signal to the TESes, defines the TES frequency bands and hosts $\mu$mux resonators. 
In addition, the detector module
provides adequate thermal sinking of TESes and readout components, forms an RF/EMI tight seal to block all RF/EMI from entering the front of the module and keeps all module-specific mechanical resonant frequencies above our likely signal sampling rate of $\sim$256\,Hz. Each detector module is composed of four major subsystems.

%The detector module needs to be able to survive for the differential contraction between materials used (e.g. silicon vs. copper) upon cooling to mK temperatures;  

1. The all-silicon optical stack consisting of a feedhorn-coupled waveguide, a waveguide interface plate (WIP), an OMT-coupled detector wafer,
and a quarter-wave backshort; a mature technological solution\cite{hubmayr12} already demonstrated in many other applications\cite{austermann12,hubmayr16,austermann18,niemack10,Henderson16,simon16}. The feedhorn profile is similar to the design for AdvACT\cite{simon16}, but modified for AliCPT-1 and streamlined fabrication.  

2. The sub-K multiplexing readout system consists of 28 66-channel microwave SQUID multiplexer chips (1,848 resonators in total) housed in a Cu packaging and wire bonded to a Routing Wafer; the latter routes the microwave signals, integrates the passive TES voltage bias circuitry, and routes the low-frequency
flux-ramp modulation required for stable $\mu$mux operation\cite{Mates11}. This independent $\mu$mux sub-assembly
is mounted on the backside of the optical stack. A combination of pin-and-slot alignment features and BeCu springs constrain and align the $\mu$mux sub-package while allowing for differential contraction relative to the all-silicon optical stack when cooled to cryogenic temperatures.
Interface printed circuit boards (PCBs) transfer the signals to standardized and accessible connectors. 
The Cu packaging creates a tightly controlled ground and field enclosures for the MUX chips using a highly conductive material. This eliminates sources of loss and helps maintain high quality factors for the resonators.
This copper sub-module solution is being developed at NIST as a general solution for all $\mu$mux projects.

3. The mechanical housing and interface to the cryostat sub-K plate: a Cu base with spring mounts for the silicon stack and an Al radiation shield. Six OFHC spring flex tabs, mounted at the vertices of the hex footprint and to indentations in the feedhorn stack, mechanically support the all-silicon optical stack and fix the x/y/z position of the feedhorn array, but with tolerance to radial differential contraction. Acting together, these 6 flex tabs are highly flexible to radially symmetric displacements (e.g. differential contraction), but extremely stiff to any translation shifts in x/y/z. 
They are pre-tensioned to the maximum expected differential contraction displacement of
0.2\,mm at room temperature, and will be at approximately 0.0\,mm displacement (and lowest stress) when cold. FEA of the stress on the tabs at maximum displacement show the maximum stress is below the yield (permanent deformation) of the material and the lowest frequency resonant mode of the structure is $>300\,$Hz.
The OFHC flex tabs, gold wire/ribbon bonds
between the main copper ring and the relevant Si wafers assure reliable cooling paths across the detector module. 
Spiral metal gasketing is used to create a flexible RF seal between the Al outer shield of the module and the conductive side walls of the silicon feedhorn array, while the waveguide section of the feedhorn prevents RF contamination via the optical path.  Thick superconducting ground planes, deposited on wafer surfaces near the sensitive components, will provide shielding from 
fluctuations in the residual magnetic fields.

%This sub-system includes flex parts allowing for thermal contraction due to differential coefficients of thermal expansion (CTE) between the metal and silicon components, all while maintaining alignment, thermal control, and stability/stiffness. 

4. A free space metal-mesh low-pass filter\cite{Ade06} is mounted at the top of the module Al shield and resides just above the feedhorn apertures.  While the primary passband forming elements are in the microwave on-chip filters of the detector array, this filter provides additional filtering of frequencies above 5.7\,icm ($\sim$171\,GHz).

The final detector passbands are formed from four primary components. 1. The waveguide acts as strong high-pass filter with $\sim$76\,GHz cutoff frequency, defining the low-frequency edge of the 90\,GHz band. 2. On-chip, quarter-wave stub filters shape the high-frequency edge of the 90\,GHz band, and both edges of the 150\,GHz band. 3. The OMT finite bandwidth, 2.25:1\cite{Datta14}, provides extra cutoff at the high-end of the 150\,GHz band. 4. Finally, the low-pass filter, in conjunction with the on-chip microwave filters, helps to
create a sharp edge to the upper (150\,GHz) bandpass, helping block a bright atmospheric emission line near $\sim$180\,GHz and high-frequency harmonic leaks of the on-chip filters.

\label{sec:det_module}

%\subsubsection{Readout Chain}\label{sec:Readout_Chain}
%\input{Readout_Chain}

%\subsubsection{Readout chain and room temperature electronics and software}\label{sec:readout_electronics}
\subsubsection{RF Readout}\label{sec:readout_electronics}
\begin{figure}
\begin{center}
\includegraphics[trim=0.cm 3.cm 0.cm 1.85cm, width=0.8\textwidth]{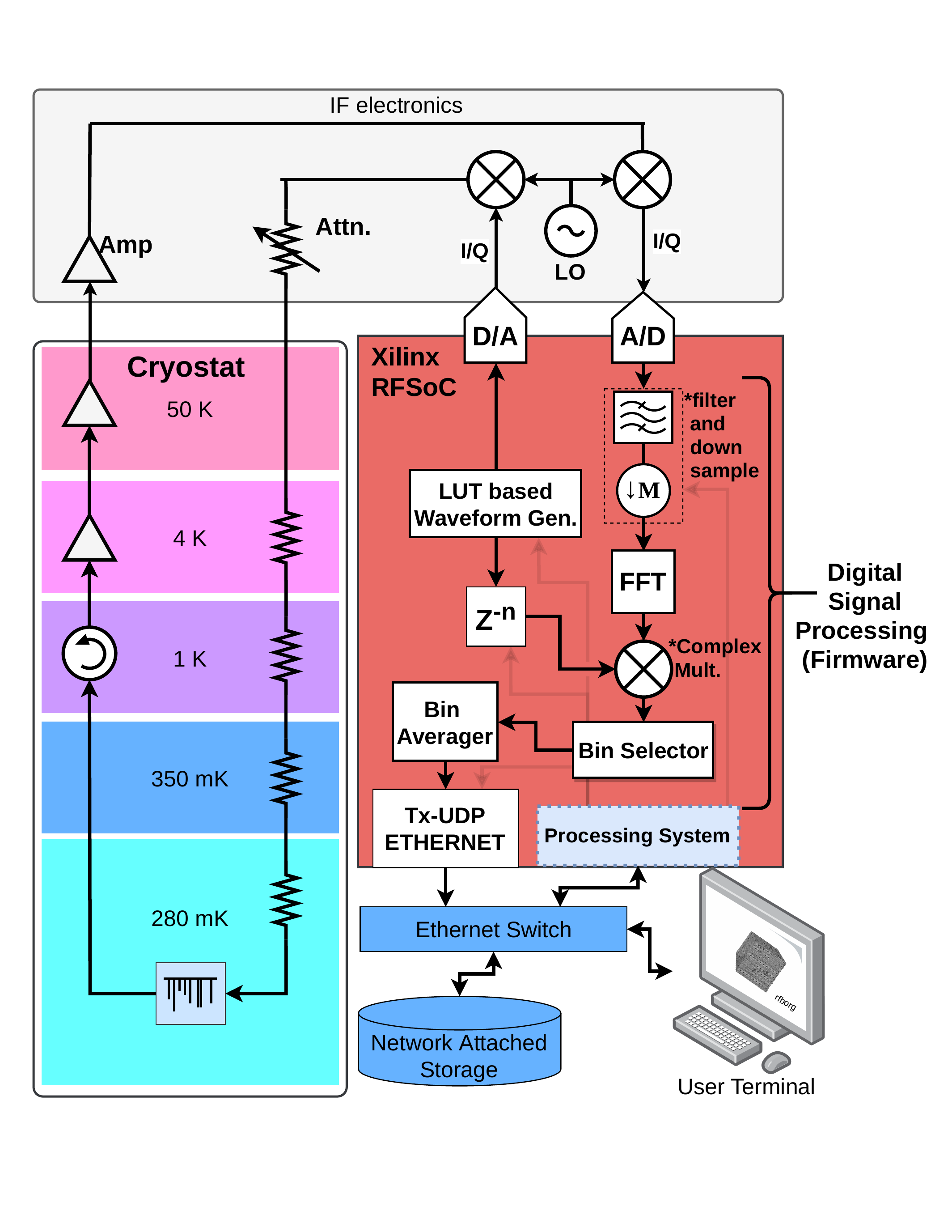}
\end{center}
\caption{Simplified readout block diagram for AliCPT-1 for a lab installation. The diagram displays the four distinct domains, cryogenic RF, intermediate frequency electronics,  high speed reconfigurable digital electronics, and Ethernet network.  At the observatory, the lower right corner ethernet-connected hardware is complex, and includes mount, HK, and site data acquisition and processing.}
\label{fig:readout_diagram}
\end{figure}

A set of RF and DC lines are required to successfully read out the $\mu$mux resonators of each detector module.  The DC lines provide bias to LNAs at 50K and 4K, along with TES bias and SQUID flux ramp to the detectors.  The RF signal requires low temperature attenuation to deliver probe tones of roughly -70 dBm to the $\mu$mux resonators.  Low noise amplifiers are then employed on the output for the tones to be read out by the room temperature electronics.  A detailed look at all the cryogenic readout components can be found in the previous AliCPT report \cite{Salatino20}. 

The multiplexed readout system is based around the ZCU111 RFSoC development  board. The RFSoC will handle the digitization and initial processing of the detector signals. A scaled up version of the BLAST-TNG firmware\cite{Gordon16} is under development base-lining a readout multiplexing (MUX) factor of 2,048 over a 2\,GHz bandwidth. To upconvert and downconvert the baseband signals to the detector frequencies we designed a transceiver  board for the 4-8\,GHz band. It features programmable attenuators and input RF amplifier and a calibration loopback trace. A Python based control software will be implemented on the Zynq microprocessor and will utilize Redis as a messaging system and in memory database. Both the firmware and software will be made open source\cite{Sinclair20}.

\subsection{Thermometry and Temperature Control}\label{sec:HK}
%%% HK subsection 

51 sensors from Lake Shore\footnote{Lake Shore Cryotronics, Inc.: \url{https://www.lakeshore.com/}.}, consisting of DT-670 diodes and Cernoxes, along with other thermometry mounted in the Chase fridge, will monitor the temperatures across different stages and locations of the cryostat, including the adsorption fridge and the FPU. Some sensors are essential for operating the receiver, others are used for monitoring and debugging purposes. 

The receiver will be provided with heaters on the 50\,K and 4\,K radiation shields to allow testing of the thermal conductance of all thermal joints. Two heaters will be used for temperature control of the FP. The adsorption fridge itself has pump and heatswitch heaters, also controlled with the HK electronics.  
Wires are routed toward three custom PCBs inside the cryostat where they will be consolidated into Manganin Tekdata cables\footnote{Tekdata Interconnections ldt,  \url{https://www.tekdata-interconnect.com/}.} and routed out of the cryostat to HK racks. The HK wires are low-pass filtered at the vacuum feedthrough using Cristek\footnote{Cristek Interconnects, Inc.:  \url{https://www.cristek.com/}.} $\pi$-type filters with 1\,MHz cutoff to minimize high frequency radio interference entering the cryostat through those wires. 

Two racks, mounted on two different sides of the VJ base side, host the warm HK electronics, 
composed of Stanford Research Systems (SRS) devices\footnote{Stanford Research Systems: \url{https://www.thinksrs.com/}.}, power supplies for the heaters.  The star camera computer will also be mounted there.  Custom software will manage the HK operation, including the fridge operation, with RS-232 serial communication to the SRS and power supply units. The HK software will be part of the control system of the receiver which manages the sky scans, detector and readout operation, and the data acquisition.

\section{Site and mount}  \label{sec:site_mount}
%%% Site and mount

\begin{figure}
\begin{center}
\includegraphics[width=1.\textwidth]{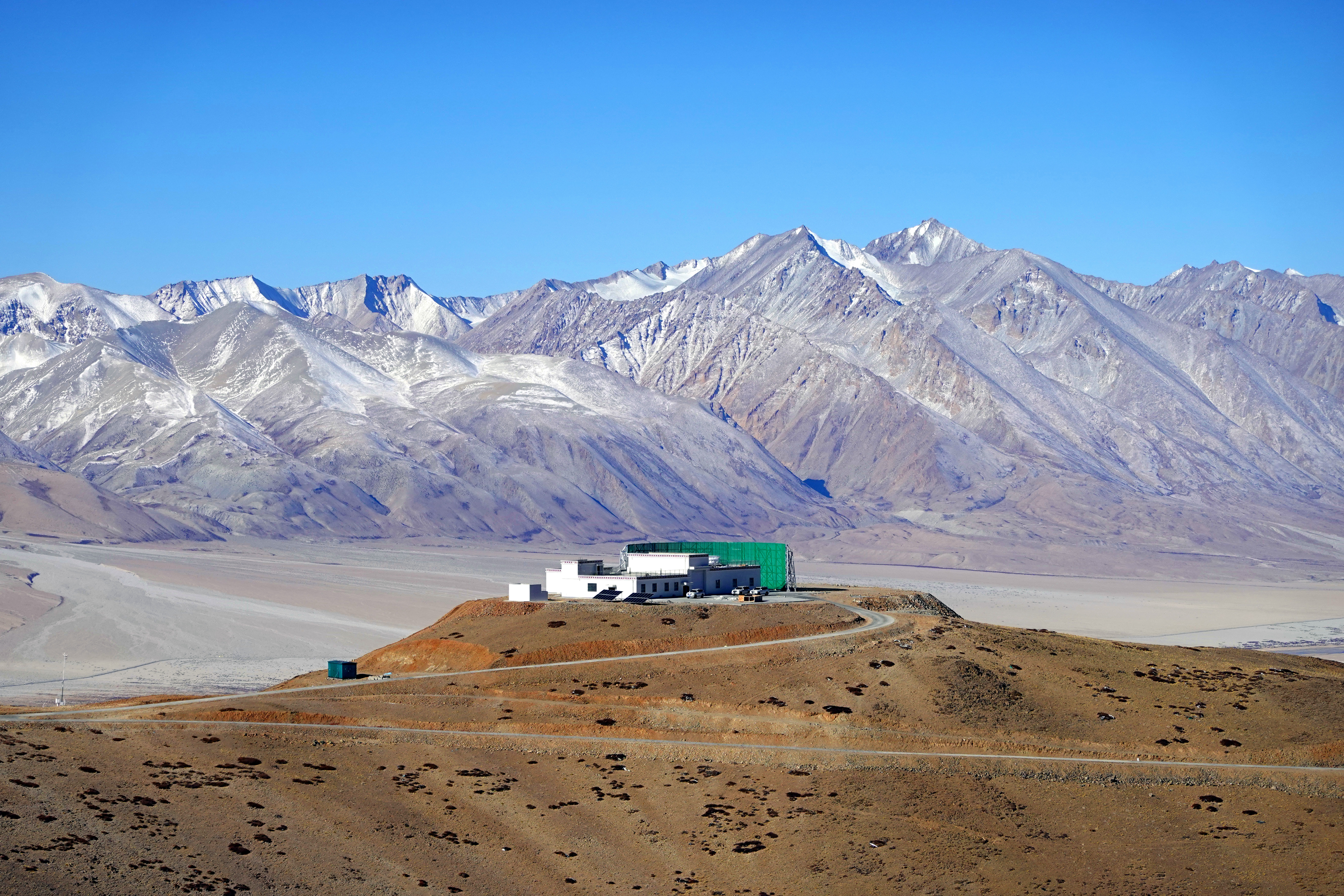}
\end{center}
\caption{The AliCPT-1 B1 site at 5,250\,m above sea level in the Ngari prefecture (Tibet).}\label{fig:site}
\end{figure}

AliCPT-1 will be deployed in the middle latitude of the Northern Hemisphere, on the Ngari (Ali) site in the Prefecture of Tibet, on a high peak of the Gangdise mountain, 32°18$^\prime$38$^{\prime\prime}$ N, 80°01$^\prime$50$^{\prime\prime}$ E at 5,250\,m above sea level (B1 site), Fig.\,\ref{fig:site}. The B1 site is located 20\,km away from the Ngari Gunsa Airport, with convenient transportation. The closest city, Shiquanhe located at 4,255m above sea level, is only 30\,km away from the B1 site.

After several years of construction, the B1 site provides excellent conditions for carrying out CMB experiments.  Infrastructure construction is complete, such as road construction, and has been connected to the city electricity power supply. AliCPT-1 will be operated from a new observatory building built by the Institute of High Energy Physics (IHEP), 850\,m$^2$  including the operation hall and additional rooms. 
The site is equipped with three weather stations monitoring pressure, wind speed and direction, and temperature.  Grid power, already ready for operation, is the main power source on site; solar panels, a diesel power generator, and a UPS power backup system are also in place. The site is also equipped with high-speed wired data service, full environmental heater/air conditioner units, and all the facilities needed to assemble and operate the receiver, including a crane and a workshop in a high-bay room. For human safety, an on-site oxygen system is present.
 
\begin{figure}
\begin{center}
\includegraphics[width=1.\textwidth]{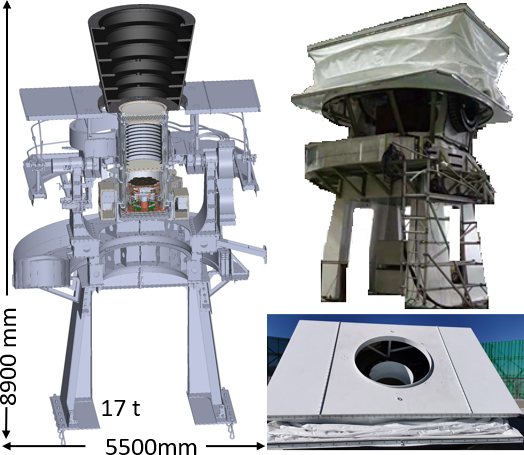}
\end{center}
\caption{Left: SolidWorks model of the receiver with the 7-detector module version forebaffle integrated on the telescope mount. Top right: the mount with the white environmental seal. Bottom right: the environmental seal and the green wind break.}\label{fig:mount}
\end{figure}

AliCPT-1 will scan the sky at high angular speed (up to 4\,$^{\circ}$/s)
to minimize the impact of the time-dependent fluctuations of atmospheric emission, which contaminate the low-frequency part of the data streams. To that effect, a fast mount (Tab.\,\ref{table:mount}, Fig.\,\ref{fig:mount}) was developed by IHEP, and has been deployed on the B1 site where it is currently being commissioned.
%AliCPT-1 will scan the sky on the mount (Tab.\,\ref{table:mount}, Fig.\,\ref{fig:mount}) developed by IHEP.
%{\color{blue}To reduce the impact of the time-dependent fluctuations of atmospheric emission, the instrument will scan the sky at high speed (up to 4$^{\circ}/$s).}
%The mount has been deployed to the B1 site and is currently being commissioned. 
A wind break has been constructed to protect the telescope from the high winds at the site. 

\begin{table}[ht]
\centering 
\begin{tabular}{|l|l|l|}
\hline
\textbf{Telescope} & Azimuth\ & $\pm270^{\circ}$\\
\textbf{Mount} & Elevation & 45°-135°{\color{blue}*} \\
& Boresight & 0°-181° \\
& Scanning speed & up to 4$^{\circ}$/s \\
\hline
\textbf{Receiver} & Dimensions & 1.4\,m x 2.5\,m  \\
         & Aperture   & 72\,cm \\
         & Weight: 7-modules     & 1,6\,t (with 150\,kg forebaffle) \\
         & Weight: 19-modules    & 2,0\,t (with 460\,kg forebaffle)\\     
\hline
\textbf{Cooling power} & PT420 & 55\,W @\,45\,K; 2.0\,W @\,4.2\,K \\
\textbf{cooling systems} & Chase fridge & 20\,$\mu$W @\,280\,mK \\
\hline 
\textbf{Signal chain} & TESes  & 32,376 90/150GHz pol-sens \\
             & $\mu$mux & current: 4-6\,GHz, 896 MUX \\
             &          & target: 4-8\,GHz, 2,048 MUX \\
\hline

\end{tabular}
\caption{AliCPT-1 synoptic overview.  *Elevation range includes motion through the zenith, allowing a boresight rotation of 180$^{\circ}$ to generate full 360$^{\circ}$ of on-sky position angle options.} 
\label{table:mount}
\end{table}

\section{Conclusion}\label{sec:conclusions}

Ground CMB observations from the Northern Hemisphere offer interesting opportunities for cross-check with other experiments and enable observing patches of the sky not accessible from the Southern Hemisphere. AliCPT-1 will be the first large focal plane CMB polarimeter to be deployed in Tibet at 5,250m above sea level. AliCPT-1 will observe the sky with 32,376 dichroic, polarization-sensitive TESes in the 90 and 150\,GHz bands. The overall
receiver, including detectors and readout, is currently under construction, integration and testing at Stanford, NIST and ASU; the mount is under commissioning on the AliCPT-1 site.

\acknowledgments
This work was supported by the Strategic Priority Research Program of the Chinese Academy of Science, CAS (Grant No. XDB23020000) and by the Sino US cooperation project (Grant No. 2016YFE0104700).

The US team is very grateful to the ASU engineer Hamdi Mani 
for the design of the 50\,K Low Noise Amplifier. The US team is very grateful to the Fermilab scientist Sara M. Simon for help in the design of the feedhorn profile.
The Stanford team is particularly grateful to David Stricker, Dorrene Ross, Michael Slack and Scott Barton for administrative and logistic support, especially during these tough times.

% References
\nocite{*}
\bibliography{main}

\begin{thebibliography}{10}

\bibitem{2011A&A...536A...1P}
{Planck Collaboration}, ``{Planck early results. I. The Planck mission}'', {\em
  \aap}~{\bf 536}, p.~A1, 2011.
\newblock {\href{https://arxiv.org/abs/1101.2022}{Astro-ph/1101.2022}}.

\bibitem{2020A&A...641A...1P}
{Planck Collaboration}, ``{Planck 2018 results. I. Overview and the
  cosmological legacy of Planck}'', {\em \aap}~{\bf 641}, p.~A1, 2020.
\newblock {\href{https://arxiv.org/abs/1807.06205}{Astro-ph/1807.06205}}.

\bibitem{2020A&A...641A...6P}
{Planck Collaboration}, ``{Planck 2018 results. VI. Cosmological parameters}'',
  {\em \aap}~{\bf 641}, p.~A6, 2020.
\newblock {\href{https://arxiv.org/abs/1807.06209}{Astro-ph/1807.06209}}.

\bibitem{1996AnPhy.246...49K}
A.~{Kosowsky}, ``{Cosmic Microwave Background polarization.}'', {\em Annals of
  Physics}~{\bf 246}(1), pp.~49--85, 1996.
\newblock {\href{https://arxiv.org/abs/astro-ph/9501045}{Astro-ph/9501045}}.

\bibitem{1997PhRvD..55.7368K}
M.~{Kamionkowski}, A.~{Kosowsky}, and A.~{Stebbins}, ``{Statistics of Cosmic
  Microwave Background polarization}'', {\em \prd}~{\bf 55}(12),
  pp.~7368--7388, 1997.
\newblock {\href{https://arxiv.org/abs/astro-ph/9611125}{Astro-ph/9611125}}.

\bibitem{1997NewA....2..323H}
W.~{Hu} and M.~{White}, ``{A CMB polarization primer}'', {\em New Astron.}~{\bf
  2}(4), pp.~323--344, 1997.
\newblock {\href{https://arxiv.org/abs/astro-ph/9706147}{Astro-ph/9706147}}.

\bibitem{1997PhRvD..55.1830Z}
M.~{Zaldarriaga} and U.~{Seljak}, ``{All-sky analysis of polarization in the
  microwave background}'', {\em \prd}~{\bf 55}(4), pp.~1830--1840, 1997.
\newblock {\href{https://arxiv.org/abs/astro-ph/9609170}{Astro-ph/9609170}}.

\bibitem{2002Natur.420..772K}
{J.M. Kovac, E.M. Leitch, C. Pryke, et \emph{al.}}, ``{{Detection of
  polarization in the Cosmic Microwave Background using DASI}}'', {\em
  Nature}~{\bf 420}(6917), pp.~772--787, 2002.
\newblock {\href{https://arxiv.org/abs/astro-ph/0209478}{Astro-ph/0209478}}.

\bibitem{2004Sci...306..836R}
{{A.C.S. Readhead}, {S.T. Myers}, {T.J. Pearson}, et \emph{al.}},
  ``{Polarization Observations with the Cosmic Background Imager}'', {\em
  Science}~{\bf 306}(5697), pp.~836--844, 2004.
\newblock {\href{https://arxiv.org/abs/astro-ph/0409569}{Astro-ph/0409569}}.

\bibitem{2006ApJ...647..813M}
{{T.E. Montroy}, {P.A.R. Ade}, {J.J. Bock}, et \emph{al.}}, ``{A Measurement of
  the CMB EE Spectrum from the 2003 Flight of BOOMERANG}'', {\em \apj}~{\bf
  647}(2), pp.~813--822, 2006.
\newblock {\href{https://arxiv.org/abs/astro-ph/0507514}{Astro-ph/0507514}}.

\bibitem{2009ApJ...692.1247P}
{{C. Pryke}, {P. Ade}, {J. Bock}, et \emph{al.}}, ``{Second and Third Season
  QUaD Cosmic Microwave Background Temperature and Polarization Power
  Spectra}'', {\em \apj}~{\bf 692}(2), pp.~1247--1270, 2009.
\newblock {\href{https://arxiv.org/abs/0805.1944}{Astro-ph/0805.1944}}.

\bibitem{2007ApJS..170..335P}
{{L. Page}, {G. Hinshaw}, {E. Komatsu}, et \emph{al.}}, ``{Three-Year Wilkinson
  Microwave Anisotropy Probe (WMAP) Observations: Polarization Analysis}'',
  {\em \apjs}~{\bf 170}(2), pp.~335--376, 2007.
\newblock {\href{https://arxiv.org/abs/astro-ph/0603450}{Astro-ph/0603450}}.

\bibitem{2009ApJ...705..978B}
{{QUaD Collaboration}: {M.L. Brown}, {P. Ade}, {J.J. Bock}, et \emph{al.}},
  ``{Improved Measurements of the Temperature and Polarization of the Cosmic
  Microwave Background from QUaD}'', {\em \apj}~{\bf 705}(1), pp.~978--999,
  2009.
\newblock {\href{https://arxiv.org/abs/0906.1003}{Astro-ph/0906.1003}}.

\bibitem{2010ApJ...711.1123C}
{{H.C. Chiang}, {P.A.R. Ade}, {D. Barkats}, et \emph{al.}}, ``{Measurement of
  Cosmic Microwave Background Polarization Power Spectra from Two Years of
  BICEP Data}'', {\em \apj}~{\bf 711}(2), pp.~1123--1140, 2010.
\newblock {\href{https://arxiv.org/abs/0906.1181}{Astro-ph/0906.1181}}.

\bibitem{2014JCAP...10..007N}
{{S. N{\ae}ss}, {M. Hasselfield}, {J. McMahon}, et \emph{al.}}, ``{The Atacama
  Cosmology Telescope: CMB polarization at 200 $<$ l $<$ 9000}'', {\em
  \jcap}~{\bf 2014}(10), p.~007, 2014.
\newblock {\href{https://arxiv.org/abs/1405.5524}{Astro-ph/1405.5524}}.

\bibitem{2015ApJ...805...36C}
{{A.T. Crites}, {J.W. Henning}, {P.A.R. Ade}, et \emph{al.}}, ``{Measurements
  of E-Mode Polarization and Temperature-E-Mode Correlation in the Cosmic
  Microwave Background from 100 Square Degrees of SPTpol Data}'', {\em
  \apj}~{\bf 805}(1), p.~36, 2015.
\newblock {\href{https://arxiv.org/abs/1411.1042}{Astro-ph/1411.1042}}.

\bibitem{1998PhRvD..58b3003Z}
M.~{Zaldarriaga} and U.~{Seljak}, ``{Gravitational lensing effect on Cosmic
  Microwave Background polarization}'', {\em \prd}~{\bf 58}(2), p.~023003,
  1998.
\newblock {\href{https://arxiv.org/abs/astro-ph/9803150}{Astro-ph/9803150}}.

\bibitem{2002ApJ...574..566H}
{W. Hu and T. Okamoto}, ``{Mass Reconstruction with Cosmic Microwave Background
  Polarization}'', {\em \apj}~{\bf 574}(2), pp.~566--574, 2002.
\newblock {\href{https://arxiv.org/abs/astro-ph/0111606}{Astro-ph/0111606}}.

\bibitem{2006PhR...429....1L}
A.~{Lewis} and A.~{Challinor}, ``{Weak gravitational lensing of the CMB}'',
  {\em Physics Report}~{\bf 429}(1), pp.~1--65, 2006.
\newblock {\href{https://arxiv.org/abs/astro-ph/0601594}{Astro-ph/0601594}}.

\bibitem{2014ApJ...794..171P}
{Polarbear Collaboration}, ``{A Measurement of the Cosmic Microwave Background
  B-mode Polarization Power Spectrum at Sub-degree Scales with POLARBEAR}'',
  {\em \apj}~{\bf 794}(2), p.~171, 2014.
\newblock {\href{https://arxiv.org/abs/1403.2369}{Astro-ph/1403.2369}}.

\bibitem{2013PhRvL.111n1301H}
{{D. Hanson}, {S. Hoover}, {A. Crites}, et \emph{al.}}, ``{Detection of B-Mode
  Polarization in the Cosmic Microwave Background with Data from the South Pole
  Telescope}'', {\em Phys. Rev. Lett.}~{\bf 111}(14), p.~141301, 2013.
\newblock {\href{https://arxiv.org/abs/1307.5830}{Astro-ph/1307.5830}}.

\bibitem{2016ApJ...833..228B}
{BICEP2 Collaboration} and {Keck Array Collaboration}, ``{BICEP2/Keck Array
  VIII: Measurement of Gravitational Lensing from Large-scale B-mode
  Polarization}'', {\em \apj}~{\bf 833}(2), p.~228, 2016.
\newblock {\href{https://arxiv.org/abs/1606.01968}{Astro-ph/1606.01968}}.

\bibitem{2016ARA&A..54..227K}
{M. Kamionkowski, E.D. Kovetz and D. Ely}, ``{The Quest for B Modes from
  Inflationary Gravitational Waves}'', {\em Annual Rev. of Astron. and
  Astroph.}~{\bf 54}, pp.~227--269, 2016.
\newblock {\href{https://arxiv.org/abs/1510.06042}{Astro-ph/1510.06042}}.

\bibitem{2018JCAP...04..016F}
{{F. Finelli}, {M. Bucher}, {A. Ach{\'u}carro}, et \emph{al.}}, ``{Exploring
  cosmic origins with CORE: Inflation}'', {\em \jcap}~{\bf 2018}(4), p.~016,
  2018.
\newblock {\href{https://arxiv.org/abs/1612.08270}{Astro-ph/1612.08270}}.

\bibitem{2020arXiv200812619T}
{The CMB-S4 Collaboration}, ``{CMB-S4: Forecasting Constraints on Primordial
  Gravitational Waves}.''
\newblock {Submitted to \apj.
  \href{https://arxiv.org/abs/2008.12619}{Astro-ph/2008.12619}, 2020}.

\bibitem{Harrington16}
{K. Harrington, T. Marriage, A. Ali et \emph{al.}}, ``{The Cosmology Large
  Angular Scale Surveyor}'', {\em Proc. of SPIE}~{\bf 9914}, p.~99141K, 2016.
\newblock {\href{https://arxiv.org/abs/1608.08234}{Astro-ph/1608.08234}}.

\bibitem{Galitzki18}
{N. Galitzki, A. Ali, K.S. Arnold et \emph{al.}}, ``{The Simons Observatory:
  instrument overview}'', {\em Proc. of SPIE}~{\bf 10708}, p.~1070804, 2018.
\newblock \href{https://arxiv.org/abs/1808.04493}{Astro-ph/1808.04493}.

\bibitem{S419}
{The CMB-S4 Collaboration}, ``{CMB-S4 Science Case, Reference Design, and
  Project Plan}'', 2019.
\newblock \href{https://arxiv.org/abs/1907.04473}{Astro-ph/1907.04473}.

\bibitem{Hamilton20}
{J.-Ch. Hamilton, L. Mousset, E.S. Battistelli, et \emph{al.}}, ``{QUBIC I:
  Overview and ScienceProgram}.''
\newblock {Submitted to \emph{JCAP}.
  \href{https://arxiv.org/abs/2011.02213}{Astro-ph/2011.02213}, 2020}.

\bibitem{2016SPIE.9914E..1JG}
{N. N. Gandilo, P.A.R. Ade, D. Benford, et \emph{al.}}, ``{The Primordial
  Inflation Polarization Explorer (PIPER)}'', {\em Proc. SPIE}~{\bf 9914},
  p.~99141J, 2016.
\newblock {\href{https://arxiv.org/abs/1607.06172}{Astro-ph/1607.06172}}.

\bibitem{2018JLTP..193.1112G}
{R. Gualtieri, J.P. Filippini, P.A.R. Ade, et \emph{al.}}, ``{SPIDER: CMB
  Polarimetry from the Edge of Space}'', {\em \jltp}~{\bf 193}(5-6),
  pp.~1112--1121, 2018.
\newblock {\href{https://arxiv.org/abs/1711.10596}{Astro-ph/1711.10596}}.

\bibitem{PdB12}
{P. de Bernardis, S. Aiola, G. Amico \emph{al.}}, ``{SWIPE: A bolometric
  polarimeter for the Large-Scale Polarization Explore}'', {\em Proc. of
  SPIE}~{\bf 8452}, p.~84523F, 2012.
\newblock \href{https://arxiv.org/abs/1208.0282}{Astro-ph/1208.0282}.

\bibitem{2011JCAP...07..025K}
{A. Kogut, D.J. {Fixsen}, D.T. {Chuss}, et \emph{al.}}, ``{The Primordial
  Inflation Explorer (PIXIE): a nulling polarimeter for cosmic microwave
  background observations}'', {\em \jcap}~{\bf 2011}(7), p.~025, 2011.
\newblock {\href{https://arxiv.org/abs/1105.2044}{Astro-ph/1105.2044}}.

\bibitem{2014JCAP...02..006A}
{Ph. Andr{\'e}, C. Baccigalupi, A. Banday, et \emph{al.}}, ``{PRISM (Polarized
  Radiation Imaging and Spectroscopy Mission): an extended white paper}'', {\em
  \jcap}~{\bf 2014}(2), p.~006, 2014.
\newblock {\href{https://arxiv.org/abs/1310.1554}{Astro-ph/1310.1554}}.

\bibitem{2018JCAP...04..014D}
{{J. Delabrouille}, {P. de Bernardis}, {F.~R. Bouchet}, et \emph{al.}},
  ``{Exploring cosmic origins with CORE: Survey requirements and mission
  design}'', {\em \jcap}~{\bf 2018}(4), p.~014, 2018.
\newblock {\href{https://arxiv.org/abs/1706.04516}{Astro-ph/1706.04516}}.

\bibitem{2019JLTP..194..443H}
{{M. Hazumi}, P.A.R. {Ade}, Y. {Akiba}, et \emph{al.}}, ``{LiteBIRD: A
  Satellite for the Studies of B-Mode Polarization and Inflation from Cosmic
  Background Radiation Detection}'', {\em \jltp}~{\bf 194}(5-6), pp.~443--452,
  2019.

\bibitem{2019arXiv190901591D}
{J. {Delabrouille}, M. H. {Abitbol}, N. {Aghanim}, et \emph{al.}}, ``{Microwave
  Spectro-Polarimetry of Matter and Radiation across Space and Time}.''
\newblock {\href{https://arxiv.org/abs/1909.01591}{Astro-ph/1909.01591}, 2019}.

\bibitem{2014PhRvL.112x1101B}
{BICEP2 Collaboration}, ``{Detection of B-Mode Polarization at Degree Angular
  Scales by BICEP2}'', {\em Phys. Rev. Lett.}~{\bf 112}(24), p.~241101, 2014.
\newblock {\href{https://arxiv.org/abs/1403.3985}{Astro-ph/1403.3985}}.

\bibitem{2015PhRvL.114j1301B}
{BICEP2/Keck Collaboration} and {Planck Collaboration}, ``{Joint Analysis of
  BICEP2/Keck Array and Planck Data}'', {\em Phys. Rev. Lett.}~{\bf 114}(10),
  p.~101301, 2015.
\newblock {\href{https://arxiv.org/abs/1502.00612}{Astro-ph/1502.00612}}.

\bibitem{BK15}
{Keck Array and BICEP2 Collaborations}, ``{BICEP2/Keck Array X: Constraints on
  Primordial Gravitational Waves using Planck, WMAP, and New BICEP2/Keck
  Observations through the 2015 Season}'', {\em Phys. Rev. Lett.}~{\bf 121},
  p.~221301, 2018.
\newblock \href{https://arxiv.org/abs/1810.05216?}{Astro-ph/1810.05216}.

\bibitem{2013ApJ...765..159D}
B.~T. {Draine} and B.~{Hensley}, ``{Magnetic Nanoparticles in the Interstellar
  Medium: Emission Spectrum and Polarization}'', {\em \apj}~{\bf 765}(2),
  p.~159, 2013.
\newblock {\href{https://arxiv.org/abs/1205.7021}{Astro-ph/1205.7021}}.

\bibitem{2013A&A...553A..96D}
{J. Delabrouille, M. Betoule, J.B. Melin et \emph{al.}}, ``{The pre-launch
  Planck Sky Model: a model of sky emission at submillimetre to centimetre
  wavelengths}'', {\em \aap}~{\bf 553}, p.~A96, 2013.
\newblock {\href{https://arxiv.org/abs/1207.3675}{Astro-ph/1207.3675}}.

\bibitem{2013JCAP...06..041M}
P.~{Mertsch} and S.~{Sarkar}, ``{Loops and spurs: the angular power spectrum of
  the Galactic synchrotron background}'', {\em \jcap}~{\bf 2013}(6), p.~041,
  2013.
\newblock {\href{https://arxiv.org/abs/1304.1078}{Astro-ph/1304.1078}}.

\bibitem{2016A&A...588A..65K}
{N. Krachmalnicoff, C. Baccigalupi, J. Aumont, et \emph{al.}},
  ``{Characterization of foreground emission on degree angular scales for CMB
  B-mode observations. Thermal dust and synchrotron signal from Planck and WMAP
  data}'', {\em \aap}~{\bf 588}, p.~A65, 2016.
\newblock {\href{https://arxiv.org/abs/1511.00532}{Astro-ph/1511.00532}}.

\bibitem{2017MNRAS.469.2821T}
B.~{Thorne}, J.~{Dunkley}, D.~{Alonso}, and S.~{N{\ae}ss}, ``{The Python Sky
  Model: software for simulating the Galactic microwave sky}'', {\em
  \mnras}~{\bf 469}(3), pp.~2821--2833, 2017.
\newblock {\href{https://arxiv.org/abs/1608.02841}{Astro-ph/1608.02841}}.

\bibitem{2017A&A...603A..62V}
{F. Vansyngel, F. Boulanger, T. Ghosh et \emph{al.}}, ``{Statistical
  simulations of the dust foreground to Cosmic Microwave Background
  polarization}'', {\em \aap}~{\bf 603}, p.~A62, 2017.
\newblock {\href{https://arxiv.org/abs/1611.02577}{Astro-ph/1611.02577}}.

\bibitem{2018MNRAS.476.1310M}
G.~{Mart{\'\i}nez-Solaeche}, A.~{Karakci}, and J.~{Delabrouille}, ``{A 3-D
  model of polarized dust emission in the Milky Way}'', {\em \mnras}~{\bf
  476}(1), pp.~1310--1330, 2018.
\newblock {\href{https://arxiv.org/abs/1706.04162}{Astro-ph/1706.04162}}.

\bibitem{2020arXiv201102221K}
N.~{Krachmalnicoff} and G.~{Puglisi}, ``{ForSE: a GAN based algorithm for
  extending CMB foreground models to sub-degree angular scales}.''
\newblock submitted to\apj.
  \href{https://arxiv.org/abs/2011.02221}{Astro-ph/2011.02221}, 2020.

\bibitem{2005MNRAS.360..935T}
M.~{Tucci}, E.~{Mart{\'\i}nez-Gonz{\'a}lez}, P.~{Vielva}, and
  J.~{Delabrouille}, ``{Limits on the detectability of the CMB B-mode
  polarization imposed by foregrounds}'', {\em \mnras}~{\bf 360}(3),
  pp.~935--949, 2005.
\newblock {\href{https://arxiv.org/abs/astro-ph/0411567}{Astro-ph/0411567}}.

\bibitem{2009A&A...503..691B}
{M. Betoule, E. Pierpaoli, J. Delabrouille, et \emph{al.}}, ``{Measuring the
  tensor to scalar ratio from CMB B-modes in the presence of foregrounds}'',
  {\em \aap}~{\bf 503}(3), pp.~691--706, 2009.
\newblock {\href{https://arxiv.org/abs/0901.1056}{Astro-ph/0901.1056}}.

\bibitem{2009AIPC.1141..222D}
{{J. Dunkley}, {A. Amblard}, {C. Baccigalupi} et \emph{al.}}, ``{Prospects for
  polarized foreground removal}'', in {\em CMB Polarization Workshop: Theory
  and Foregrounds: CMBPol Mission Concept Study},  {S. Dodelson et \emph{al.}},
  ed., {\em American Institute of Physics Conference Series} {\bf 1141},
  pp.~222--264, 2009.
\newblock {\href{https://arxiv.org/abs/0811.3915}{Astro-ph/0811.3915}}.

\bibitem{2016MNRAS.458.2032R}
{M. Remazeilles, C.Dickinson, et \emph{al.}}, ``{Sensitivity and foreground
  modelling for large-scale Cosmic Microwave Background B-mode polarization
  satellite missions}'', {\em \mnras}~{\bf 458}(2), pp.~2032--2050, 2016.
\newblock {\href{https://arxiv.org/abs/1509.04714}{Astro-ph/1509.04714}}.

\bibitem{2017PhRvD..95d3504A}
D.~{Alonso}, J.~{Dunkley}, B.~{Thorne}, and S.~{N{\ae}ss}, ``{Simulated
  forecasts for primordial B-mode searches in ground-based experiments}'', {\em
  \prd}~{\bf 95}(4), p.~043504, 2017.
\newblock \href{https://arxiv.org/abs/1608.00551}{Astro-ph/1608.00551}.

\bibitem{2018JCAP...04..023R}
{{M. Remazeilles}, {A.J. Banday}, {C. Baccigalupi}, et \emph{al.}},
  ``{Exploring cosmic origins with CORE: B-mode component separation}'', {\em
  \jcap}~{\bf 2018}(4), p.~023, 2018.
\newblock {\href{https://arxiv.org/abs/1704.04501}{Astro-ph/1704.04501}}.

\bibitem{2018ApJ...853..127H}
B.~S.~{Hensley} and P.~{Bull}, ``{Mitigating Complex Dust Foregrounds in Future
  Cosmic Microwave Background Polarization Experiments}'', {\em \apj}~{\bf
  853}(2), p.~127, 2018.
\newblock {\href{https://arxiv.org/abs/1709.07897}{Astro-ph/1709.07897}}.

\bibitem{2019PhRvD..99d3529E}
{J. Errard and R. Stompor}, ``{Characterizing bias on large scale CMB B-modes
  after Galactic foregrounds cleaning}'', {\em \prd}~{\bf 99}(4), p.~043529,
  2019.
\newblock {\href{https://arxiv.org/abs/1811.00479}{Astro-ph/1811.00479}}.

\bibitem{2009LNP...665..159D}
J.~{Delabrouille} and J.-F. {Cardoso}, {\em {Diffuse Source Separation in CMB
  Observations}}, vol.~665, pp.~159--205.
\newblock {Editor: V.J. Mart{\'\i}nez et \emph{al.}}, 2009.
\newblock {\href{https://arxiv.org/abs/astro-ph/0702198}{Astro-ph/0702198}}.

\bibitem{2015ApJ...809...63E}
{{J. Errard}, {P.A.R. Ade}, {Y. Akiba}, et \emph{al.}}, ``{Modeling Atmospheric
  Emission for CMB Ground-based Observations}'', {\em \apj}~{\bf 809}(1),
  p.~63, 2015.
\newblock {\href{https://arxiv.org/abs/1501.07911}{Astro-ph/1501.07911}}.

\bibitem{benson14}
{B.A. Benson, P.A.R. Ade, Z. Ahmed, et \emph{al.}}, ``{SPT-3G: a
  next-generation Cosmic Microwave Background polarization experiment on the
  South Pole telescope}'', {\em Proc. SPIE}~{\bf 9153}, p.~91531P, 2014.
\newblock {\href{https://arxiv.org/abs/1407.2973}{Astro-ph/1407.2973}}.

\bibitem{Ahmed14}
{Z. Ahmed, M. Amiri, S.J. Benton, et. \emph{al}}, ``{BICEP3: a 95 GHz
  refracting telescope for degree-scale CMB polarization}'', {\em Proc.
  SPIE}~{\bf 9153}, p.~91531N, 2014.
\newblock \href{https://arxiv.org/abs/1407.5928}{Astro-ph/1407.5928}.

\bibitem{Schillaci20}
{A. Schillaci, P.A.R. Ade, Z. Ahmed et \emph{al.}}, ``{Design and Performance
  of the First BICEP Array Receiver}'', {\em JLTP}~{\bf 199}, p.~976–984,
  2020.
\newblock \href{https://arxiv.org/abs/2002.05228}{Astro-ph/2002.05228}.

\bibitem{Wu09}
{J.-H.P. Wu, P.T.P. Ho, C.-W.L. Huang et \emph{al.}}, ``{Array for Microwave
  Background Anisotropy: Observations, Data Analysis, and Results for
  Sunyaev-Zel'Dovich Effects}'', {\em ApJ}~{\bf 694}(2), pp.~1619--1628, 2009.

\bibitem{2012SPIE.8444E..2YR}
{{J.A. Rubi{\~n}o-Mart{\'\i}n}, {R. Rebolo}, {M. Aguiar} et \emph{al.}}, ``{The
  QUIJOTE-CMB experiment: studying the polarisation of the galactic and
  cosmological microwave emissions}'', {\em Proc. SPIE} {\bf 8444}, p.~84442Y,
  2012.

\bibitem{2014JLTP..176..787M}
{A. Monfardini, R. Adam, R. Adane, et \emph{al.}}, ``{Latest NIKA Results and
  the NIKA-2 Project}'', {\em \jltp}~{\bf 176}(5-6), pp.~787--795, 2014.
\newblock {\href{https://arxiv.org/abs/1310.1230}{Astro-ph/1310.1230}}.

\bibitem{Merra2}
``Merra-2 website.''
\newblock \url{https://gmao.gsfc.nasa.gov/reanalysis/MERRA-2/}.

\bibitem{Kuo17}
{C.-L. Kuo}, ``{Assessments of Ali, Dome A, and Summit Camp for mm-wave
  Observations Using MERRA-2 Reanalysis}'', {\em ApJ}~{\bf 848}(1), p.~64,
  2017.
\newblock \href{https://arxiv.org/abs/1707.08400}{Astro-ph/1707.08400}.

\bibitem{LiYP17}
{Y.-P. Li, Y. Liu, S.-Y. Li, et \emph{al.}}, ``{Tibet's Ali: A New Window to
  Detect the CMB Polarization}.''
\newblock \href{https://arxiv.org/abs/1709.09053}{Astro-ph/1709.09053}, 2017.

\bibitem{LiH18}
{H. Li, S.-Y. Li, Y. Liu, et \emph{al.}}, ``{Tibet's window on primordial
  gravitational waves}'', {\em Nature Astron.}~{\bf 2}, pp.~104--106, 2018.
\newblock \href{https://arxiv.org/abs/1802.08455}{Astro-ph/1802.08455}.

\bibitem{LiH19}
{H. Li, S.-Y. Li, Y. Liu, et \emph{al.}}, ``{Probing Primordial Gravitational
  Waves: Ali CMB Polarization Telescope}'', {\em National Science Review}~{\bf
  6}(1), p.~145–154, 2019.
\newblock {\href{https://arxiv.org/abs/1710.03047}{Astro-ph/1710.03047}}.

\bibitem{Lapinov20}
{V. Lapinov, S.A. Lapinova, L.Y. Petrov and D. Ferrusca}, ``{On the benefits of
  the Eastern Pamirs for sub-mm astronomy}'', {\em Proc. of SPIE}~{\bf 11453},
  p.~114532O, 2020.
\newblock \href{https://arxiv.org/abs/2012.04647}{Astro-ph/2012.04647}.

\bibitem{GMAO}
``{NASA GEOS FP-IT data website}.''
\newblock \url{https://gmao.gsfc.nasa.gov/GMAO_products/NRT_products.php}.

\bibitem{2019BAAS...51g..57L}
{{M. Levi}, {L.E. Allen}, {A. Raichoor} et \emph{al.}}, ``{The Dark Energy
  Spectroscopic Instrument (DESI)}'', in {\em Bulletin of the American
  Astronomical Society},   {\bf 51}, p.~57, 2019.
\newblock {\href{https://arxiv.org/abs/1907.10688}{Astro-ph/1907.10688}}.

\bibitem{Minami19}
{Y. Minami, H. Ochi, K. Ichiki, et \emph{al.}}, ``{Simultaneous determination
  of the cosmic birefringence and miscalibrated polarisation angles from CMB
  experiments}'', {\em Progress of Theoretical and Experimental Physics}~{\bf
  2019}(8), p.~083E02, 2019.
\newblock {\href{https://arxiv.org/abs/1904.12440}{Astro-ph/1904.12440}}.

\bibitem{SYLi15}
{S.-Y. Li, J.-Q. Xia, M. Li, et \emph{al.}}, ``{Testing CPT Symmetry with
  Current and Future CMB Measurements}'', {\em \apj}~{\bf 799}(2), p.~211,
  2015.
\newblock {\href{https://arxiv.org/abs/1405.5637}{Astro-ph/1405.5637}}.

\bibitem{Choi13}
{{J. Choi, H. Ishitsuka, S. Mima, et \emph{al.}}}, ``{Radio-transparent
  multi-layer insulation for radiowave receivers}'', {\em Review of Scientific
  Instruments}~{\bf 84}, 2013.

\bibitem{Kang18}
{J. H. Kang, et. \emph{al}}, ``{2017 upgrade and performance of BICEP3: a 95GHz
  refracting telescope for degree-scale CMB polarization}'', {\em Proc.
  SPIE}~{\bf 10708}, p.~107082N, 2018.
\newblock \href{https://arxiv.org/abs/1808.00567}{Astro-ph/1808.00567}.

\bibitem{Inoue14}
{Y. Inoue, T. Matsumura, M. Hazumi, et \emph{al.}}, ``{Cryogenic infrared
  filter made of alumina for use at millimeter wavelength}'', {\em Applied
  Optics}~{\bf 53}, pp.~1727--1733, 2014.
\newblock {\href{https://arxiv.org/abs/1311.5388}{Astro-ph/1311.5388}}.

\bibitem{Raguin_Morris93}
{D.H. Raguin and G.M. Morris}, ``{Analysis of antireflection-structured
  surfaces with continuous one-dimensional surface profiles}'', {\em Applied
  Optics}~{\bf 32}(14), pp.~2582--2598, 1993.

\bibitem{Biber03}
{S. Biber, J. Richter, S. Martius, and L. Schmidt}, ``{Design of Artificial
  Dielectrics for Anti-Reflection-Coatings}'', {\em {Proceedings of the 33rd
  European Microwave Conference}}~{\bf 3}, pp.~1115--1118, 2003.
\newblock Munich.

\bibitem{Salatino20}
{M. Salatino, J. Austermann, J. Meinke, A.K. Sinclair, S. Walker et
  \emph{al.}}, ``{Current status of the Ali CMB Polarization Telescope Focal
  Plane camera}.''
\newblock Submitted to \emph{IEEE Trans. on Appl. Supercond.}, Nov 2020.

\bibitem{NISTdatabase}
``{NIST} cryogenic material properties database.''
\newblock \url{https://trc.nist.gov/cryogenics/index.html}.

\bibitem{RunyanJones2008}
{M.C. Runyan and W.C. Jones}, ``{Thermal conductivity of thermally-isolating
  polymeric and composite structural support materials between 0.3 and 4\,K}'',
  {\em Cryogenics}~{\bf 48}, pp.~448--454, Sep 2008.
\newblock {\href{https://arxiv.org/abs/0806.1921}{Astro-ph/0806.1921}}.

\bibitem{Irwin04}
{K.D. Irwin and K.W. Lehnert}, ``{Microwave SQUID multiplexer}'', {\em
  Appl.Phys. Lett.}~{\bf 85}, p.~2107, 2004.

\bibitem{Mates11}
{J.A.B. Mates}, {\em {The Microwave SQUID Multiplexer}}.
\newblock PhD thesis, University of Colorado, 2011.

\bibitem{Dober17}
{B. Dober, D.T. Becker, D.A. Bennett et \emph{al.}}, ``{Microwave SQUID
  Multiplexer demonstration for Cosmic Microwave Background Imagers}'', {\em
  Appl.Phys. Lett.}~{\bf 111}, p.~243510, 2017.
\newblock \href{https://arxiv.org/abs/1710.04326}{Astro-ph/1710.04326}.

\bibitem{duff16}
{S.M. Duff, J. Austermann, J.A. Beall, et \emph{al.}}, ``{Advanced ACTPol
  multichroic polarimeter array fabrication process for 150\,mm wafers}'', {\em
  \jltp}~{\bf 184}(3-4), pp.~634--641, 2016.

\bibitem{hubmayr16}
{J. Hubmayr, J.E. Austermann, J.A. Beall, et \emph{al.}}, ``{Design of 280 GHz
  feedhorn-coupled TES arrays for the balloon-borne polarimeter SPIDER}'', {\em
  Proc. of SPIE}~{\bf 9914}, p.~99140V, 2016.
\newblock {\href{https://arxiv.org/abs/1606.09396}{Astro-ph/1606.09396}}.

\bibitem{walker20a}
{S. Walker, C.E. Sierra, J.E. Austermann, et \emph{al.}}, ``{Demonstration of
  220/280 GHz Multichroic Feedhorn-Coupled TES Polarimeter}'', {\em \jltp}~{\bf
  199}, p.~891–897, 2020.
\newblock {\href{https://arxiv.org/abs/1909.11569}{Astro-ph/1909.11569}}.

\bibitem{Walker20}
{S. Walker, J.E. Austermann, J.A. Beall et \emph{al.}}, ``{Measurements of AlMn
  TES bolometers well described by simple electrothermal model}'', {\em Proc.
  of SPIE}~{\bf 11453}, p.~1145326, 2020.

\bibitem{hubmayr12}
{J. Hubmayr, J.W., J.E. Austermann, et \emph{al.}}, ``{An all silicon
  feedhorn-coupled focal plane for Cosmic Microwave Background polarimetry}'',
  {\em \jltp}~{\bf 167}(5-6), pp.~904--910, 2012.

\bibitem{austermann12}
{J.E. Austermann, K.A. Aird, J.A. Beall,\emph{al.}}, ``{SPTpol: an instrument
  for CMB polarization measurements with the South Pole Telescope}'', {\em
  Proc. of SPIE}~{\bf 8452}, p.~84521E, 2012.

\bibitem{austermann18}
{J.E. Austermann, J.A. Beall, S.A. Bryan, \emph{al.}}, ``{Millimeter-wave
  polarimeters using kinetic inductance detectors for TolTEC and beyond}'',
  {\em \jltp}~{\bf 193}(3), pp.~120--127, 2018.
\newblock {\href{https://arxiv.org/abs/1803.03280}{Astro-ph/1803.03280}}.

\bibitem{niemack10}
{M.D. Niemack, P.A. Ade, J. Aguirre, et \emph{al.}}, ``{ACTPol: a
  polarization-sensitive receiver for the Atacama Cosmology Telescope}'',  {\bf
  7741}, p.~77411S, 2010.
\newblock {\href{https://arxiv.org/abs/1006.5049}{Astro-ph/1006.5049}}.

\bibitem{Henderson16}
{S.W. Henderson, R. Allison, {J. Austermann et \emph{al.}}}, ``{Advanced
  {ACTP}ol Cryogenic Detector Arrays and Readout}'', {\em \jltp}~{\bf
  184}(3-4), pp.~772--779, 2016.
\newblock \href{https://arxiv.org/abs/1510.02809}{Astro-ph/1510.02809}.

\bibitem{simon16}
{S.M. Simon, J. Austermann, J.A. Beall et \emph{al.}}, ``{The design and
  characterization of wideband spline-profiled feedhorns for Advanced
  ACTPol}'', {\em Proc. SPIE}~{\bf 9914}, p.~991416, 2016.

\bibitem{Ade06}
{P.A.R. Ade, G. Pisano, C. Tucker, and S. Weaver}, ``{A review of metal mesh
  filters}'', {\em Proc. SPIE}~{\bf 6275}, p.~62750U, 2006.

\bibitem{Datta14}
{R. Datta, J. Hubmayr, C. Munson et \emph{al.}}, ``{Horn coupled multichroic
  polarimeters for the Atacama Cosmology Telescope polarization experiment}'',
  {\em \jltp}~{\bf 176}, pp.~1670--676, 2014.
\newblock {\href{https://arxiv.org/abs/1401.8029}{Astro-ph/1401.8029}}.

\bibitem{Gordon16}
{S. Gordon, B. Dober, A. Sinclair, et \emph{al.}}, ``{An Open Source,
  FPGA-Based LeKID Readout for BLAST-TNG: Pre-Flight Results}'', {\em Journal
  of Astronomical Instrumentation}~{\bf 5}, p.~1641003, 2020.
\newblock {\href{https://arxiv.org/abs/1611.05400}{Astro-ph/1611.05400}}.

\bibitem{Sinclair20}
{A.K. Sinclair, R.C. Stephenson, J. Hoh et \emph{al.}}, ``{On the development
  of a reconfigurable readout for superconducting arrays}'', {\em Proc. of
  SPIE}~{\bf 11453}, p.~114531T, 2020.

\end{thebibliography}
\bibliographystyle{spiebib2} % makes bibtex use spiebib.bst

\end{document}